\RequirePackage{fix-cm}
\documentclass[twocolumn,epjc3]{svjour3}  
\smartqed  
\RequirePackage{graphicx}

\RequirePackage{mathptmx}      
\RequirePackage{flushend}
\RequirePackage[numbers,sort&compress]{natbib}
\RequirePackage[colorlinks,citecolor=blue,urlcolor=blue,linkcolor=blue]{hyperref}
\RequirePackage{multirow}
\usepackage{epstopdf}
\usepackage{subfig}
\usepackage{comment}
\usepackage{float}

\journalname{Eur. Phys. J. ST}
\begin{document}


\title{Charmonium production in low energy nuclear collisions at SPS and FAIR: achievements $\&$ prospects}



\author{Partha Pratim Bhaduri\thanksref{e1,addr1, addr2}
}

\thankstext{e1}{e-mail: partha.bhaduri@vecc.gov.in}


\institute{Variable Energy Cyclotron Centre, 1/AF Bidhannagar, Kolkata - 700064 \label{addr1}
           \and
           Homi Bhabha National Institute, Anushaktinagar, Mumbai \label{addr2}
}

\date{Received: date / Accepted: date}

\maketitle

\begin{abstract}
In this article, we review the status of the charmonium production in low energy fixed target proton-nucleus (p-A) and nucleus-nucleus (A-A) collisions as measured by different experimental collaborations at CERN-SPS, Fermilab and HERA accelerator facilities. The interplay of different cold and hot medium effects influencing the production of these $c\bar{c}$ bound states at low collision energies is discussed in detail. Prospect for upcoming charmonium measurements close to kinematic production threshold, in the CBM experiment at FAIR SIS100 and NA60+ experiment at CERN-SPS facilities are also investigated.

\end{abstract}

\section{Introduction}
\label{intro}

\begin{table*}
\begin{center}
\begin{tabular}{|c|c|c|c|c|c|}
\hline
  Facility  & Experiment & Mode & Max. interaction rate (Hz) & $\sqrt{s_{NN}}$(GeV) & $\mu_{B}$ (MeV)   \\
  \hline
   SIS18  & HADES & Fixed Target & 2 $\times 10^{4}$ & 2.4 - 2.6 & 800 - 770 \\
  \hline
    HIAF  & CEE & Fixed Target & 5 $\times 10^{5}$ & 2 - 2.7 & 880 - 770 \\ 
  \hline
    Nuclotron  & BM@N & Collider & 5 $\times 10^{4}$  & 2 - 3.5 & 880 - 670 \\ 
  \hline
  J-PARC-HI  & DHS, D2S & Collider &  $10^{7}$ & 2 - 6.2 & 880 - 430 \\ 
  \hline
  SIS100  & CBM & Fixed Target & $10^{7}$ &2.7 - 4.9 & 760 - 500 \\ 
  \hline
   NICA  & MPD & Collider & $4 \times 10^{3}$  &4 - 11 & 580 - 300 \\ 
  \hline
   RHIC  & STAR-BES & Fixed Target $\&$ Collider & $10^{2}$ - $10^{3}$  &3 - 19.6 & 720 - 210 \\ 
  \hline
 SPS  & NA61 & Fixed Target & $10^{3}$   & 4.9 - 17.3 & 520 - 230 \\ 
  \hline
  SPS  & NA60+ & Fixed Target & $1.5 \times 10^{5}$ & 4.9 - 17.3 & 520 - 230 \\ 
  \hline
\end{tabular}
\caption{Experimental programs either in operation or under construction or being planned to probe the high $\mu_{B}$ region of the QCD phase diagram. The accelerator facility, the experiment, mode of operation, maximum interaction rate, $\sqrt{s_{NN}}$ and $\mu_B$ coverage are listed.}
\label{tab:facility}
\end{center}
\end{table*}

Relativistic heavy-ion collision experiments offer us the unique opportunity to study the formation and characterization of the quark-gluon plasma (QGP) in the laboratory~\cite{Bhalerao:2014owa,Andronic:2014zha,Braun-Munzinger:2015hba,Elfner:2022iae,Harris:2023tti}. By varying the collision energy, hot and dense strongly interacting matter can be produced over a wide range of temperatures and net baryon densities which in principle help us to probe the QCD phase diagram  and search for the possible transitions between the confined hadronic matter phase and the deconfined quark-gluon matter~\cite{Satz:2018oiz}. Heavy-ion collisions at ultra-relativistic energies like those reached at the (top) RHIC and the LHC have probed the hot QCD matter at high temperature ($T$) and small baryo-chemical potential ($\mu_{B}$) regime. State of the art lattice QCD simulations have established a smooth cross over transition for vanishingly small $\mu_{B}$ at a pseudo-critical temperature of $T_{pc} = 156.5 \pm 1.5 $ MeV~\cite{HotQCD:2018pds,HotQCD:2019xnw}. Measurements carried out at these two major accelerator facilities indicate that the produced partonic medium is very strongly coupled, characterized by transport coefficients having magnitudes very close to conjectured lower bounds obtained in strongly coupled quantum field theories based on gauge/gravity duality (Ads/CFT). Detailed characterization of these transport properties and their connection to the quark-hadron phase transition, remain as central questions which are being addressed with upgrade programs at LHC energies~\cite{ALICE:2022wpn}. As compared to high $T$ and low $\mu_{B}$ region, till date the QCD phase structure is much less explored in the region of large baryon densities. Highly compressed nuclear matter at relatively low temperatures is likely to exist in the Universe, inside the core of the neutron stars and their mergers. Theoretical model calculations suggest that with increasing $\mu_{B}$, the smooth cross over transition between the hadronic and the QGP phase, eventually becomes a discontinuous first order phase transition through a second order critical end point (CEP)~\cite{Fukushima:2010bq,Luo:2022mtp}. Lattice QCD calculations do not find a critical end point till ${\mu_{B} \over T} \le 3.0$~\cite{HotQCD:2019xnw}. The so called fermionic sign problem~\cite{Gattringer:2016kco} prohibits the application of lattice QCD in the large $\mu_{B}$ region. Recent theoretical calculations suggest that the location of CEP in the QCD phase diagram is largely driven by the microscopic degrees of freedom of QCD, quarks and gluons~\cite{Gunkel:2021oya}. Functional methods of QCD, based on Dyson–Schwinger equations (DSE)~\cite{Gunkel:2021oya,Gao:2020fbl} and/or the functional renormalisation group (fRG)~\cite{Fu:2019hdw,Fu:2023lcm}, enable to map out the entire QCD phase diagram and constrain the location of the CEP to $600 < \mu_{B} < 650$ MeV. The current fRG approach with a QCD-assisted low-energy effective theory (LEFT) finds the locus of the CEP to be at $(T_{CEP},\mu_{B_{CEP}}) = (98,643)$ MeV~\cite{Fu:2023lcm}. In the analysis of the trajectory of the Lee-Yang edge singularities of the QCD equation of state in the complex $\mu_{B}$ plane, the CEP is 
predicted at $(T_{CEP},\mu_{B_{CEP}}) = (100, 580)$ MeV~\cite{Basar:2023nkp}. For $\mu_{B} \ge 417$ MeV indications of an inhomogeneous region in the vicinity and above the chiral transition are also seen and a consistent picture of the QCD phase structure at finite baryon densities has begun to emerge.  To complement the theoretical studies, over the past two decades there have been enormous efforts, to experimentally scan the moderate $T$ and high $\mu_{B}$ regime of the QCD phase diagram, by systematically lowering the collision energy. Dedicated experiments at different accelerator facilities around the globe, are either taking data, undergoing construction or being planned for investigation of the properties of dense nuclear matter in the high $\mu_{B}$ regime~\cite{Galatyuk:2019lcf}. Table~\ref{tab:facility} provides a summary of such experimental programs including their mode of operation, maximum achievable interaction rate (for heavy-ion collisions), centre-of-mass energy and the corresponding range of $\mu_{B}$. \\

One of the major objectives of any relativistic heavy-ion collision program is to find unambiguous and experimentally viable observables suitable to characterize the hot and dense fireball produced in such reactions. Bound states made of charm ($c$) and anti-charm ($\bar{c}$) quarks which do not decay via strong interaction are collectively called charmonia~\cite{Eichten:1978tg}. The $J/\psi$ meson, the ground state vector boson, is the most widely studied resonance in charmonium family. It was discovered in 1974 simultaneously at BNL in 23 GeV p+Be collisions by the E598 Collaboration~\cite{E598:1974sol} and at SLAC in $e^+e^-$ collisions~\cite{SLAC-SP-017:1974ind}. Even after intense theoretical and experimental investigations over last 50 years since the $J/\psi$ state was discovered, a complete understanding of the charmonium production in hadronic collisions that gives a coherent description of all the experimental observables is still missing~\cite{Brambilla:2010cs,Bodwin:2012ft,Chen:2021tmf}. The production involves three different momentum scales, namely the charm quark mass ($m_{c} \simeq 1.5$ GeV) governing the perturbative production of the $c\bar{c}$ pairs, the momentum $m_{c}v$ of the heavy quark in the charmonium rest frame ($v$ being the charm quark velocity) and the typical charm quark kinetic energy $m_{c}v^{2}$ that drives the non-perturbative hadronization of the evolving $c\bar{c}$ pairs to physical resonance states. Due to the small values of velocity ($v^{2} \simeq 0.3$), charmonia are usually treated as non-relativistic bound states. In literature, charmonium production in elementary nucleon-nucleon collisions is commonly treated via a factorizable two component model. Creation of the $c\bar{c}$ pairs in hard collisions with momentum transfer, $Q^{2} \ge 4m_{c}^{2}$, constitutes the first step and is described by perturbative QCD (pQCD). At high collision energies, the dominant contribution to the total charm production cross section comes from gluon gluon fusion ($gg \rightarrow c\bar{c}$), whereas quark anti-quark annihilation ($q\bar{q} \rightarrow c\bar{c}$) also significantly contributes to the inclusive $c\bar{c}$ pair production at lower energies. Subsequently in the second stage, the $c\bar{c}$ pairs initially produced in a color octet state evolve to form the color neutral physical resonances (eg: $J/\psi$, $\psi(2S)$ or $\chi_{c}$) with appropriate size and specific quantum numbers. The process of color neutralization by interacting with the adjoining color field and the subsequent formation of physical
resonance is presumably soft and non-perturbative in nature. Different models are available 
in literature to describe the hadronization of the $c\bar{c}$ pairs to color neutral bound states~\cite{Schuler:1994hy,Gavai:1994in,Schuler:1996ku,Lansberg:2002cz}. One simple and experimentally well supported phenomenological approach is the color evaporation model (CEM)~\cite{Fritzsch:1977ay,Gluck:1977zm}, where the $c$ and the $\bar{c}$ can either get 
combined with the more abundant light up ($u$) and down ($d$) quarks resulting in the formation of open charm hadrons (eg: $D^{\pm}$, $\Lambda_{c}$ etc.). Alternatively they may bind with each other forming a charmonium state, during the evaporation process. In this formulation, the basic quantity is the so called closed charm production cross section, $\Sigma_{c\bar{c}}$, calculated by integration of the perturbative $c\bar{c}$ production cross section $\sigma$ over the pair mass window between $2m_{C}$ and $2m_{D}$. 
According to CEM, for the {\it i}-th charmonium state the production  cross section becomes a constant fraction of the closed charm cross section, $\sigma_{i} (s) = f_{i} \Sigma_{c\bar{c}} (s)$, $f_{i}$ being an energy independent empirical constant. Thus the energy dependence of the production cross section of any charmonium state is fully determined by the pQCD estimated closed charm cross section. Although CEM has been found successful to describe $J/\psi$ production in hadronic collisions over a wide energy range, it has a very serious flaw. A direct outcome of the CEM approach is that the relative yields of various charmonia states are independent of the kinematics and independent of the colliding species. In contrast, the experimentally measured $\psi(2S)/J/\psi$ ratio has been found to depend on transverse momentum ($p_T$). Recently an improved version of the color evaporation model (ICEM)~\cite{Ma:2016exq} has been proposed in literature. By distinguishing between exchanged and emitted soft gluons and imposing  constraints on the invariant mass and momentum shifts between the intermediate heavy quark pair ($Q\bar{Q}$) and the physical resonance state, ICEM can nicely reproduce the transverse momentum ($p_T$) dependence of the $\psi(2S)/J/\psi$ ratio.  The other leading approaches include color singlet model (CSM)~\cite{Ellis:1976fj,Carlson:1976cd} and non-relativistic QCD (NRQCD)~\cite{Bodwin:1994jh,Brambilla:1999xf} factorization approach among the others. 
 The main assumption of CSM is that the initially produced heavy quark pairs take the same orbital angular momentum ($L$), spin ($S$) and colour quantum numbers of the physical quarkonium state since they are produced in hard partonic collisions and no orbital ($L$) or spin ($S$) angular-momentum-changing transition takes place during resonance binding. Subsequent measurements of $J/\psi$ and $\psi(2S)$ production by experiments at Fermilab showed that the model calculations severely underestimate the data~\cite{DK2097,DK2098}. To overcome the significantly large discrepancy between model calculations and measured data, the NRQCD factorization approach was introduced. The theoretical construct of NRQCD framework includes heavy quark pair production in both color-octet and color-singlet channels. The total cross-section is a sum of the contributions from these two different states, weighted by long-distance matrix elements (LDMEs), which are introduced as
 model parameters converting the pre-resonant coloured $Q\bar{Q}$ pairs into physical bound states, possibly changing  $L$ and/or $S$ as required. These parameters are universal (i.e. process-independent) and constant (i.e. independent of the $Q\bar{Q}$ momentum) and their values are fixed by comparison to experimental data. The production cross section of different quarkonium states as computed within the NRQCD factorization approach has been seen to be in reasonable agreement with the experimental data recorded at Tevatron~\cite{DK2018,DK2019,DK2020}, 
 RHIC~\cite{DK2015,DK2016,DK2017} and LHC~\cite{DK2007,DK2008,DK2009,DK2010,DK2011,DK2012,DK2013,DK2014}. A comprehensive review of different quarkonium production models can be found at~\cite{Brambilla:2010cs,QuarkoniumWorkingGroup:2004kpm,Andronic:2015wma}. \\

\begin{table*}
\begin{tabular}{|c|c|c|c|c|c|c|}
\hline
  Experiment   & Facility & Collision System & $E_{beam}$(GeV) & Targets  & phase space   \\
  \hline
   NA38   & SPS & p+A & 450 & Be,Al,Cu,Ag,W,Pb &$-0.425<y_{cms}<0.575$ \\
  \hline
   NA38   & SPS & p+A & 200 & Be,Al,Cu,Ag,W,Pb &$-0.425<y_{cms}<0.575$ \\
  \hline
   NA38   & SPS & O+Cu, O+U, S+U & 200 A & Be,Al,Cu,Ag,W,Pb & $-0.425<y_{cms}<0.575$ \\
  \hline
   NA51   & SPS & p+p,p+d & 400 & Be,Al,Cu,Ag,W,Pb & $-0.425<y_{cms}<0.575$ \\
  \hline
  NA50   & SPS & p+A & 400 & Be,Al,Cu,Ag,W,Pb & $-0.425<y_{cms}<0.575$ \\
  \hline
  NA50  & SPS & p+A & 450 & Be,Al,Cu,Ag,W & $-0.5<y_{cms}<0.5$ \\
  \hline 
  NA50  & SPS & Pb+Pb & 158 & Pb &  $0.0<y_{cms}<1.0$ \\
  \hline 
  NA60 & SPS & p+A & 400& Be,Al,Cu,Ag,W,Pb,U & $-0.17<y_{cms}<0.33$ \\
  \hline
  NA60 & SPS & p+A & 158 & Be,Al,Cu,Ag,W,Pb,U & $0.28<y_{cms}<0.78$ \\
  \hline
  NA60 & SPS & In+In & 158 & In  & $0.0<y_{cms}<1.0$ \\
  \hline
  E866 & Tevatron & p+A & 800& Be,Fe,W & $-0.6<y_{cms}<2.46$ \\
  \hline
  E906 & Tevatron & p+A & 120 & C,Fe,W & $0.86<y_{cms}<1.47$ \\
  \hline
  HERA-B & HERA & p+A & 920 & C,Ti,W & $-2.2<y_{cms}<1.8$ \\
  \hline
\end{tabular}
\caption{Basic features of the (selected) fixed target  experiments measuring charmonium ($J/\psi$ $\&$ $\psi(2S)$) production in proton and ion induced collisions.}
\label{tab:Expt. details}
\end{table*}

In spite of our limited knowledge about the production process, charmonia are considered as a useful diagnostic probe to study the hot and dense nuclear matter produced in high energy heavy-ion collisions~\cite{Vogt:1999cu,Kluberg:2009wc,Andronic:2014sga,Arnaldi:2014aca,Arnaldi:2015sra,Scomparin:2016gog,Arnaldi:2016cjt,Arnaldi:2017stq,Bhaduri:2020yea,Arnaldi:2020rep,Arnaldi:2021idt,Andronic:2025jbp}. The suppression of the yield of charmonium states in relativistic nuclear collisions is one of the early signatures proposed for studying the de-confinement transition and QGP formation in the laboratory. In 1986, Matsui $\&$ Satz~\cite{Matsui:1986dk}, prior to any experimental observation, predicted that inside a color deconfined QGP medium the Debye screening of gluons and quarks would prevent the resonant binding of $c$, $\bar{c}$ pairs leading to reduced production of $J/\psi$ mesons as compared to a scenario where a deconfined medium is absent. The screening radius, $\lambda_{D}$ which characterizes the maximum separation between $c$ and $\bar{c}$ in a bound state that survives dissociation, depends on the system temperature, with stronger screening in hotter plasma. Since the different $c\bar{c}$ bound states have considerably different sizes and binding energies, they undergo dissociation at different temperatures, with loosely bound states being melted first. This leads to a sequential suppression pattern with increasing energy density of the medium and can, in principle, be used to estimate the temperature of the medium~\cite{Digal:2001ue,Karsch:2005nk}. In hadronic collisions, a substantial fraction of the observed $J/\psi$ mesons does not come from direct production, but originates from the decay of heavier excited charmonium states, about $8 \%$ from the hadronic decay of $\psi(2S)$ state and $22 \%$ from the radiative decay of $\chi_{c1,2}(P)$ states. In heavy-ion collisions, the melting of these loosely bound excited states in the QGP occurs at lower temperatures than those required to dissociate the tightly bound $J/\psi$ states. This suppresses their survival probability and ultimately reduces the number of $J/\psi$ mesons produced via decay. The suppression of the $J/\psi$ yield, observed in nuclear collisions can be partly attributed to the reduced feed down due to the melting of its excited states inside the plasma, even if the  the medium is not hot enough to directly dissolve the strongly bound $1S$ ground states.\\

Since the seminal work by Matsui and Satz, considerable efforts have been invested over the last 40 years, to develop state-of-the-art models to describe the effects of the hot and dense deconfined medium on quarkonium production. It is well known by now that the in-medium quarkonium potential is complex in nature. 
Considerable theoretical efforts have been invested to study the behavior of the finite temperature $Q\bar{Q}$ potential. A recent investigation based on lattice QCD inputs combined with effective field theory methods has revealed that the real part of the potential changes very little with temperature of the medium~\cite{Bazavov:2023dci}. This in turn indicates that the $Q\bar{Q}$ force is essentially not affected by Debye screening, in contrast to the original charmonium melting picture proposed by Matsui and Satz. Instead the imaginary part of the potential grows significantly with distance and temperature, leading to large in-medium widths and rapid dissociation of the quarkonium states. Similar conclusions are reached in a recent study of in-medium $Q\bar{Q}$ potential within hard thermal loop (HTL) approach~\cite{Duari:2025kxt}. Hence as per our contemporary understanding, quarkonium suppression is driven not by weakened binding, as in the original paradigm, but by dynamical in-medium dissociation process. \\

The two major theoretical formulations, in vogue, for description of quarkonium evolution in QGP are: (i) the statistical hadronization model (SHM) and (ii) transport models based on semiclassical kinetic rate equations and/or semiclassical Boltzmann equations or open quantum systems framework. The statistical hadronization model for charm (SHMc)~\cite{Braun-Munzinger:2000csl,Andronic:2006ky,Andronic:2008gm,Andronic:2021erx} is based on the assumption that the charm quarks are produced in the initial hard collisions and that their number remains conserved until hadronization. Due to Debye screening in the QGP medium all the primordial charmonium states undergo complete dissociation. The corresponding charm quarks undergo 
complete thermalization in the expanding medium and recombine again at the QCD phase boundary along with the hadronization of light quarks and gluons. At high collision energies, the chemical freeze-out temperature ($T_{ch}$) of the expanding hot QCD medium, estimated from light-flavor hadron yields, coincides with the temperature of the QCD crossover transition from lattice QCD calculations. Thus information of the inclusive $c\bar{c}$ cross section along with $T_{ch}$ obtained from the analysis of the yields of light hadrons is sufficient to determine the total yield of all hadrons containing charm quarks within a grand canonical ensemble framework. Assuming kinetic freeze-out of the heavy hadrons to coincide with their chemical freeze-out at the QCD phase boundary and employing hydrodynamics, SHMc can also reproduce the transverse momentum ($p_{T}$) spectra of the $J/\psi$ mesons at the LHC.
\\

In the semiclassical transport approach~\cite{Thews:2000rj,Grandchamp:2001pf,Wu:2024gil,Rapp:2017chc,Du:2022uvj}, the time evolution of the instantaneous yield $N(\tau)$ ($\tau$ being the time in the QGP frame) of a given charmonium state is simultaneously dictated by a loss term due to suppression of the initially produced charmonia and a production term corresponding to the regeneration of charmonia due to the recombination of the $c\bar{c}$ pairs in the fireball. The $c$ and $\bar{c}$ quarks undergoing recombination may come from the endogamous i.e. same initial $c\bar{c}$ pair ("diagonal term") or from exogamous i.e. two distinct primordial pairs ("off-diagonal term"). In this dynamical approach, unlike SHM approach, both the dissociation and regeneration processes remain operational throughout the life time of the plasma. The time evolution of the fireball temperature, $T(\tau)$, is tuned through hydrodynamical model calculations. The current implementations of the kinetic approach vary in their model inputs for the temperature dependence of the reaction rates ($\Gamma(T)$) and also for the in-medium binding energies of the different quarkonium states. In recent years transport approaches based on open quantum system (OQS) frameworks have been developed to provide a rigorous non-unitary framework for modeling quarkonium evolution in the QGP~\cite{Akamatsu:2020ypb,Yao:2021lus,Brambilla:2024tqg,Delorme:2024rdo}. The OQS framework treats the $Q\bar{Q}$ pairs together with the QGP as a closed quantum system represented by the density matrix, $\rho_{tot}$, evolving unitarily. By tracing out the degrees of freedom of the $Q\bar{Q}$ states, one ends up with non-unitary and time-irreversible evolution equations, the quantum master equations, for the reduced density matrix, $\rho_{Q\bar{Q}}$. In certain limits, mainly when the in medium binding energy of a specific quarkonium state is much smaller compared to the fireball, the master equations can be reduced to the so-called Lindblad equation which is a non-unitary generalization of the Schrödinger equation that unifies screening, dissociation, and recombination processes inside a QGP. A comprehensive discussion on the features of different state-of-the-art theoretical models and their inter-comparison can be found in Ref.~\cite{Andronic:2024oxz}. \\

A considerable amount of charmonium suppression in nuclear collisions may also originate from several effects not related to a deconfined medium. In particular, a sizable suppression is also observed in proton-nucleus (p+A) collisions. In such collisions, the formation of an extended QGP medium is not traditionally expected. \footnote{However there is an ongoing debate on the possibility of the formation of a small QGP droplet in small collision systems. Data collected in the last 10 years in p+Pb and high multiplicity p+p collisions at the LHC exhibit some features like collective flow, strangeness enhancement, which in heavy-ion collisions are understood as due to QGP formation~\cite{ALICE:2022wpn,Grosse-Oetringhaus:2024bwr}}. The observed charmonium suppression in p+A collisions is attributed to the various initial and final state effects of distinct physics origins, operative on different stages of resonance formation in cold nuclear matter (CNM).  Various CNM effects as identified in literature include nuclear modification of the parton distribution functions~\cite{Eskola:1998df,Vogt:2004dh,deFlorian:2003qf,Eskola:2009uj}, Cronin effect~\cite{Cronin:1974zm,NA38:1991rhp}, break-up of the initially-produced $c\bar{c}$ pairs in the pre-resonance or resonance stage~\cite{Gerschel:1988wn} and initial/final state parton energy loss~\cite{Vogt:1999dw,Arleo:2012hn}. Precise estimation of In such collisions the  level of the "normal" nuclear suppression of the charmonium states due to CNM effects in A+A collisions is important to isolate the plausible "anomalous" suppression due to genuine hot medium effects in the heavy-ion data. 
This is based on the assumption that no QGP is formed in p+A collisions and charmonium production is affected only by cold nuclear matter effects. Measuring charmonium production in p+A collisions therefore helps to calibrate a robust reference baseline to compare with the suppression pattern seen in nuclear collisions. \\

The first experimental investigation of charmonium ($J/\psi$ and $\psi(2S)$) production, in nuclear collisions was carried out, immediately after the theoretical prediction by Matsui and Satz, at CERN SPS by the NA38 Collaboration~\cite{NA38:1989okg,NA38:1990fnn,NA38:1991vno} in O+U, S+U collisions at a beam energy of 200 A GeV in the laboratory frame. 
These measurements were subsequently followed by the NA50 Collaboration in Pb+Pb collisions~\cite{NA50:1997rtj,NA50:1997hlx,NA50:1999stx,NA50:2000brc,NA50:2001vsq,NA50:2004sgj} and by the NA60 Collaboration in In+In collisions~\cite{NA60:2006ncq}, both with a beam energy of 158 A GeV in the laboratory frame and within the same kinematic domain. In central Pb+Pb collisions as measured by the NA50 Collaboration, $J/\psi$ mesons were observed to be "anomalously suppressed" well beyond the dissociation induced by the CNM effects. This, together with the other measurements performed at the SPS, led CERN to announce the first evidence of QGP formation in Pb+Pb collisions in the year 2000~\cite{NA50:2000brc,Kluberg:2005yh}. The yield of $\psi(2S)$ mesons as measured in Pb+Pb collisions was found to be more reduced as compared to $J/\psi$ mesons, indicating sequential suppression pattern~\cite{Santos:2004zz,NA50:2006yzz}. The CERN SPS measurements were then followed by complementary measurements at higher centre of mass energies first at RHIC at a maximum energy of $\sqrt{s_{NN}} = 200$ GeV Au+Au collisions by the PHENIX and STAR Collaborations and later on at the CERN LHC by the ALICE, CMS and ATLAS experiments. Beginning in 2010 with $\sqrt{s_{NN}} = 2.76 $ TeV Pb+Pb collisions in the LHC Run 1 campaign, the charmonium (as well as bottomonium, the $b\bar{b}$ bound states) states are now measured at $\sqrt{s_{NN}} = 5.02$ TeV Run 2 and very recently in $\sqrt{s_{NN}} = 5.36$ TeV Pb+Pb interactions in Run 3. With the increase of collision energy, new mechanisms of charmonium production come into play. The primordial suppression of the charmonium yields, as compared to their production in p+p collisions, appears to be (partially) compensated via the regeneration process. The increasing abundance of $c\bar{c}$ quark pairs in a deconfined medium at higher collision energies enhances the probability of forming charmonia from exogamous (non-diagonal) recombination. SHMc calculations predict that in central Pb+Pb collisions, the average number of heavy quark pairs, $N_{c\bar{c}}$ is around 0.2 at top SPS energy ($\sqrt{s_{NN}} = 17.3$ GeV), about 10 at RHIC ($\sqrt{s_{NN}} = 200$ GeV) and increases up to 120 at top LHC energy ($\sqrt{s_{NN}} = 5.02$ TeV)~\cite{Andronic:2006ky}. Thus at RHIC and LHC energies, the $c\bar{c}$ pairs, either produced endogamously in the same hard collision or exogamously if $c$ and $\bar{c}$ quarks come from different collisions but close in phase space, may undergo recombination to form a charmonium state, either in the plasma phase and/or at the phase boundary when the system undergoes hadronization. This secondary production leads to an increase of the charmonium yields in more central collisions, where a larger number of $c\bar{c}$ pairs is produced in the initial hard scatterings. One may note that at lower collision energies, charmonium can, in principle, be formed via recombination of $c$ and $\bar{c}$ quarks at a (late) stage of the QGP evolution, or at the hadronization. However, with only 0.2 $c\bar{c}$ pairs produced per event in central Pb+Pb collisions at top SPS energy (meaning that, on average, one pair is produced every five collisions) only endogamous recombination is possible. In this case, the $c$ and $\bar{c}$ quarks can initially form a (pre-resonant) bound state that can dissociate in the QGP, but there is a non-zero probability that these quarks may combine together at  a later stage, forming a charmonium. Exogamous production is, instead, effectively forbidden at low energies, thus making the overall probability of charmonium regeneration very low in collisions at SPS and FAIR.
While measurements at CERN SPS were limited only to charmonium ($J/\psi$ and $\psi(2S)$) states, the increase in $\sqrt{s_{NN}}$ up to the LHC energies opens up the possibility to study the production of bottomonium states complementing the charmonium results. Quarkonium measurements are being performed by all running experiments at RHIC and LHC with different rapidity ($y$) and transverse momentum ($p_{T}$) coverage. The availability of data for different quarkonium states over an extended kinematic range and broad range of collision energies allows detailed investigation of quarkonium properties in a deconfined medium. Below we present a brief summary of the main results on charmonium production obtained in heavy-ion collisions at RHIC and LHC energies. Even though the focus of our article is on lower energies, nevertheless, some of the conclusions drawn from RHIC and LHC measurements are also relevant for a more solid interpretation of the measurements at lower energies. Moreover, data from RHIC and LHC have served as important benchmarks to constrain the previously mentioned theoretical models that may be applied or extended to lower collision energies. Finally, the RHIC Beam Energy Scan (BES) covered an energy range similar to that of SPS. Hence we also include the major highlights of charmonium production from RHIC BES studies that are recently becoming available.  \\

The most widely studied observable to investigate quarkonium production in heavy-ion interactions at RHIC and LHC is the so-called nuclear modification factor, $R_{AA}$\footnote{$R_{AA}^{J/\psi}$ is defined as $R_{AA}^{J/\psi} = {{dN_{J/\psi}^{AA}/dy} \over {<N_{coll}>.dN_{J/\psi}^{pp}/dy}}$ and relates the $J/\psi$ yield in A+A collisions to that expected for a superposition of independent p+p collisions. $<N_{coll}>$ denotes the average number of binary collisions, for a given centrality class. Due to unavailability of p+p collisions, the $J/\psi$ production in heavy-ion collisions at SPS were published in comparison to the Drell-Yan production, an electromagnetic process not affected by QGP formation.}. 
At RHIC, both PHENIX and STAR Collaborations reported a strong suppression of $J/\psi$ production in $\sqrt{s_{NN}} = 200$ GeV in central Au+Au collisions as compared to binary scaling of p+p collisions~\cite{PHENIX:2006gsi,STAR:2013eve}. The Cu+Cu data were found to be consistent with no suppression, although the precision was limited by the available statistics~\cite{STAR:2013eve}. The magnitude of suppression in Au+Au collisions was found to be somewhat weaker compared to the level of suppression detected in Pb+Pb collisions at SPS\footnote{For a comparable mean charged particle multiplicity density, $<dN_{ch}/d\eta>$, used to mimic the collision centrality.}, even though the temperature and energy density achieved at these two collision energies are significantly different. Moreover the suppression was seen to be stronger at forward rapidity ($1.2 < |y| < 2.2$) compared to mid-rapidity ($|y| < 0.35$)~\cite{PHENIX:2006gsi}. 
The combined effects of color screening of the $c\bar{c}$ potential and the subsequent regeneration via $c\bar{c}$ recombination are thought to determine the differences observed in the $J/\psi$ suppression patterns at SPS and RHIC. At RHIC, where $\sqrt{s_{NN}}$ is more than 10 times higher than at SPS, the larger charm quark multiplicity enhances regeneration, which partly compensates for the increased suppression.
This could also explain the stronger suppression at forward rapidity at RHIC since the charm quark density is larger at mid-rapidity, leading to a higher probability of forming charmonia via recombination. To further investigate the interplay of the two competing mechanisms of color screening and regeneration, $J/\psi$ production at mid-rapidity ($|y| < 1$) was measured by the STAR Collaboration in Au+Au collisions at lower energies, namely $\sqrt{s_{NN}}=39, 62.4$ GeV~\cite{STAR:2016utm}. No significant energy dependence of $J/\psi$ $R_{AA}$ was observed within uncertainties. A similar level of $J/\psi$ suppression as that at $\sqrt{s_{NN}}=200$ GeV was measured by the PHENIX Collaboration at forward rapidity ($1.2 < |y| < 2.2$) in Au+Au collisions at $\sqrt{s_{NN}} =39$ and 62.4 GeV~\cite{PHENIX:2012xtg}.
At top RHIC energy, new $J/\psi$ data were recorded by STAR at mid-rapidity for Au+Au collisions with improved precision and extended kinematic reach compared to previous measurements~\cite{STAR:2025imj}. At low $p_{T}$, an increasing suppression is observed from peripheral to central collisions due to the interplay of the CNM effects, dissociation, and $c\bar{c}$ recombination. For $p_{T} > 5$ GeV/$c$, a significant suppression of $J/\psi$ yield was observed in central collisions, most likely due to the color-screening in the deconfined medium. The $J/\psi$ $R_{AA}$ exhibited little dependence on $p_{T}$ in all centrality bins. 
Recently the STAR Collaboration has also reported preliminary results for the centrality and $p_{T}$ dependence of $J/\psi$ $R_{AA}$ in Au+Au collisions at $\sqrt{s_{NN}} =14.6, 17.3, 19.6$ and $27$ GeV at mid-rapidity ($|y|< 1$) by analyzing data collected during the RHIC Beam Energy Scan-II (BES-II) program~\cite{STAR:2025imj}. At all collision energies, a stronger suppression of the $p_{T}$ integrated $J/\psi$ yield is observed towards more central collisions, which follows a similar trend to the 200 GeV results. The centrality integrated $R_{AA}(p_{T})$ exhibits a clear $p_{T}$ dependence, with strongest suppression at low $p_{T}$ and an increasing trend with increasing $p_{T}$. CNM effects might be partially responsible for low $p_{T}$ suppression, since they are expected to be stronger with decreasing collision energy below 200 GeV. The $p_{T}$ dependence at low collision energies is steeper than the one measured at 200 GeV suggesting that a not-negligible regeneration contribution at 200 GeV more effectively compensates the suppression at low $p_{T}$. The interpretation of these RHIC results is not yet conclusive due to the presence of both suppression and (moderate) regeneration effects and also due to the complex interplay between the hot and cold matter effects. 
Additionally, the STAR Collaboration has reported the experimental observation of sequential charmonium suppression, by measuring inclusive $J/\psi$ and $\psi(2S)$ production at mid-rapidity ($|y|<1$) in $\sqrt{s_{NN}} = 200$ GeV Ru+Ru and Zr+Zr collisions~\cite{Zhang:2025jie}. The $p_{T}$-integrated double ratio, defined as the $\psi(2S)$ to $J/\psi$ ratio in heavy-ion collisions relative to that in p+p collisions
has been found to be below unity by $5.6\sigma$ in ($0 - 80 \%$) central collisions, providing evidence for stronger suppression of $\psi(2S)$ than $J/\psi$ in ion-ion collisions compared to p+p collisions. The double ratio is also smaller than the interpolated value for $\sqrt{s_{NN}} = 200$ GeV p+Au collisions, indicating additional $\psi(2S)$ suppression relative to $J/\psi$ beyond the CNM effects~\cite{STAR:2021zvb}. Hints of more pronounced sequential suppression at low $p_{T}$ are also seen. \\

The charmonium measurements in Pb+Pb collisions at the LHC, due to their very broad rapidity coverage down to very low transverse momentum, were expected to give a conclusive picture on the role played by the (re)generation mechanism on the observed production dynamics. At the LHC, the $J/\psi$ $R_{AA}$ has been measured in Pb+Pb collisions at $\sqrt{s_{NN}} = 2.76$ and $5.02$ TeV~\cite{ALICE:2012jsl,ATLAS:2010xzb,CMS:2012bms,ALICE:2016flj,ALICE:2023hou}. 
The $J/\psi$ production is seen to be significantly less suppressed at LHC energies as compared to Au+Au collisions at $\sqrt{s_{NN}} =200$ GeV measured at RHIC. These results suggest that with increasing $\sqrt{s_{NN}}$, regeneration through $c\bar{c}$ recombination becomes more important at LHC energies.
In contrast to the measurements at SPS and RHIC, the $J/\psi$ $R_{AA}$ measured at LHC exhibits a rather weak dependence on collision centrality.
The $p_{T}$ dependence of $J/\psi$ $R_{AA}$ evaluated for the most central collisions at mid-rapidity exhibits a significant $p_{T}$ dependence with stronger suppression at high $p_{T}$, in stark contrast to the corresponding observation at RHIC. Measurements from different experiments at mid-rapidity, namely ALICE ($|y|<0.9$), CMS ($|y| < 2.4$) and ATLAS ($|y| < 2$) show a very good agreement in the overlapping $p_{T}$ range. 
The thermal distribution of charm quarks peaks at low $p_{T}$ (and at $y =0$) resulting in stronger $c\bar{c}$ recombination and hence more regenerated $J/\psi$ mesons at low $p_{T}$ than at high $p_{T}$. The ALICE experiment is the most suitable facility to investigate the role of (re)generation in Pb+Pb collisions at LHC energies, by measuring charmonium production down to $p_{T} \simeq 0$\footnote{A very low $p_{T}$ cut ($p_{T} > 0.3$ GeV/$c$ is applied for rejection of $J/\psi$ mesons produced through photo-production process, which significantly contribute to the $J/\psi$ yield particularly in most peripheral collisions~\cite{ALICE:2019lga,ALICE:2021gpt}}, both at mid-rapidity ($|y| <0.9$) and at forward rapidity ($2.5 < y <4$), via di-electron ($e^{+}e^{-}$) and di-muon ($\mu^{+}\mu^{-}$) channels respectively. 
The $J/\psi$ $R_{AA}$ is higher at midrapidity with respect to forward rapidity at low $p_{T}$ ($p_{T} < 3$ GeV/c)~\cite{ALICE:2023gco}.
In addition to the extensive studies on the $J/\psi$ suppression pattern, further insights on the charmonium dynamics in the nuclear collisions can be inferred by measuring the elliptic flow ($v_{2}$) of the $J/\psi$ mesons, from the anisotropy of their azimuthal distributions. At RHIC, the $J/\psi$ $v_{2}$ was found to be compatible with zero, with almost no $p_{T}$ dependence, within large uncertainties~\cite{Qiu:2012zz,PHENIX:2024axj}. At LHC, $J/\psi$ mesons are found to have a non-zero $v_{2}$ increasing with $p_{T}$ and reaching a maximum ($v_{2}^{max} \simeq 0.1$) in semi-central collisions and at $p_{T} \sim 4$ GeV/$c$~\cite{ALICE:2013xna,ALICE:2017quqa,CMS:2023mtk}. The observed behavior also finds a natural explanation in the regeneration scenario, where a large fraction of the detected $J/\psi$ mesons are assumed to originate from the recombination of the thermalized charm quarks which acquire the medium anisotropy by participating \textcolor{red}{in} the collective expansion of the fireball. The loosely bound $\psi(2S)$ states exhibit similar features to the $J/\psi$. However the degree of suppression for $\psi(2S)$ resonance is found to be about a factor of two larger than that of $J/\psi$ over the entire $p_{T}$ or centrality range, in Pb+Pb collisions at LHC~\cite{ALICE:2022jeh}. We refrain from discussing further on quarkonium production in high energy nuclear collisions. For a comprehensive review of the recent experimental results on quarkonium production at RHIC and LHC energies, the reader may consult Ref.~\cite{Andronic:2025jbp}.



\begin{figure*}
  \includegraphics[width=1.0\linewidth]{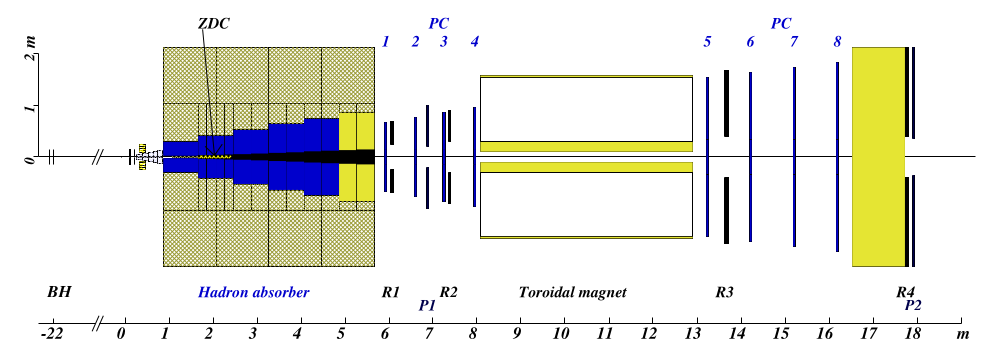}
\caption{Schematic of the muon spectrometer set up employed by \textcolor{red}{the} NA50 experiment for di-muon measurements at CERN-SPS facility~\cite{Kluberg:2005yh}.}
\label{fig:setup_NA50}       
\end{figure*}

As compared to the high energy frontier, data on quarkonium production is rather scarce in low energy nuclear collisions. Till date quarkonium production in relativistic nuclear collisions has not been explored below top SPS energy. The primary reason behind the non-availability of such measurements is the extremely small charm production cross sections at lower energies which demands accelerators with extremely high beam intensities and detectors with very good rate capability to enable such measurements. Among the different ongoing and upcoming high $\mu_{B}$ facilities, the NA60+/DiCE experiment~\cite{Scomparin:2021xvy,Borysova:2022nyr,NA60_plus_LOI,Arnaldi:2023zlh,Scomparin:2023bbu,Alocco:2024hvm,Usai:2024row,NA60DiCE:2025qra,Arnaldi:2025ikz,NA60DiCE:2025abc} at CERN SPS and Compressed Baryonic Matter (CBM) experiment~\cite{CBM:2016kpk,Agarwal:2023otg,Hohne:2024kpd} at FAIR aim at performing a detailed measurement of charmonium production in p+A and A+A collisions below $\sqrt{s_{NN}} = 17.3$ GeV. The NA60+/DiCE experiment aims to extend the existing measurements on charmonium production at SPS below 158 A GeV, down to a beam energy of 40 A GeV ($\sqrt{s_{NN}}=8.8$ GeV) or even lower depending on the integrated luminosity available at the SPS facility. The CBM experiment plans to explore near threshold or even sub-threshold charmonium production in heavy-ion collisions. Both the experiments have dedicated plans to study charmonium production in p+A collisions to complement the corresponding heavy-ion measurements. In this article, we briefly review the prospects of performing such high rate measurements in relativistic nuclear collisions. The remaining of the article is organized as follows. In section~\ref{sec:2} we summarize the main results on charmonium production in fixed target p+A and A+A collisions with emphasis on the measurements performed at SPS. The physics potential of the planned charmonium measurements in the upcoming fixed target experiments is discussed in section~\ref{sec:3}. Results from physics performance simulations for $J/\psi$ measurements using optimized detector setups at NA60+ and CBM are briefly discussed in section~\ref{sec:4}. Finally we summarize in section~\ref{sec:5}.

\section{Experimental results on charmonium suppression at fixed target facilities}
\label{sec:2}
\begin{figure}
  \includegraphics[width=1.0\linewidth]{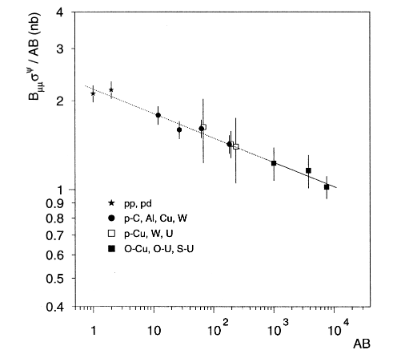}
\caption{ $J/\psi$ production cross-sections per nucleon-nucleon collisions, in the di-muon decay channel ($B_{\mu\mu} \simeq 6 \%$) plotted as a function the product of the mass numbers (A $\times$ B) of the projectile and target nuclei, measured by the NA38 and NA51 experiments at CERN SPS facility~\cite{NA38:1998udo,Abreu:1999nn}. The 450 GeV data points were rescaled to 200 GeV and reevaluated in the corresponding kinematic region.}
\label{fig:jpsi_na38}       
\end{figure}

Over the years a wealth of charmonium data, particularly focusing on $J/\psi$ production have been collected in several fixed target p+A and A+A collision experiments covering different kinematic range and energy domains, as summarized in Table~\ref{tab:Expt. details}. At CERN SPS facility, charmonium production was measured by the NA38, NA51, NA50 and NA60 experiments, for various systems starting from p+p and p+A collisions to S+U and finally in Pb+Pb and In+In collisions. The $J/\psi$ and $\psi(2S)$ mesons were detected via their di-muon ($\mu^{+}\mu^{-}$) decay channel. The NA38 experimental apparatus was based on a muon spectrometer inherited from the NA10 experiment, originally designed for the detection of thermal di-muons and a Pb-scintillating fibre electromagnetic calorimeter (EMCal). A 4.8 m thick hadron absorber made of high density graphite separated the spectrometer from the target region. A tungsten-uranium central plug was placed inside the graphite block to absorb the non-interacting beam particles. An air core magnet (ACM) of hexagonal symmetry produces the toroidal magnetic field that bends the charged particle tracks with a deflection inversely proportional to their transverse momentum. The trigger was provided by the four plastic scintillator hodoscopes (R1 - R4) and the two sets of multi wire proportional counters (MWPC), located upstream and downstream of the magnet were used for reconstruction of muon track candidates. Di-muons were reconstructed in the laboratory rapidity range $2.8 < y < 4.0$, with di-muon mass resolution of $\sim 110$ MeV/$c^2$ in the $J/\psi$ mass region. Multiple scattering inside the hadron absorber and the ACM field controlled the single track momentum resolution which in turn determined the pair mass resolution. The EMCal placed downstream of the target region in the pseudo-rapidity ($\eta$) range $1.7 < \eta < 4.1$ was employed to determine the collision centrality, by measuring the transverse energy ($E_T$) mostly deposited by neutral pions, on an event by event basis. For the subsequent NA50 experiment, the design of the muon spectrometer was kept practically unchanged. The thick hadron absorber facilitated the experiment to be operated with a highly luminous incoming beam and a moderate spectrometer illumination. However the target region was upgraded to cope with the increased multiplicity and radiation levels of Pb+Pb interactions as compared to O and S induced collisions studied by the NA38 Collaboration. The $\eta$ coverage of the EMCal detector was modified to $1.1 < \eta < 2.3$. In addition, two new detectors were also added to the apparatus: a vertex detector made of silicon strips to measure the charged particle multiplicity ($dN_{ch}/d\eta$) and a zero degree calorimeter (ZDC) housed inside the carbon absorber for measuring the energy deposited by spectator nucleons ($E_{ZDC}$) for each Pb+Pb interaction. The experimental setup used by the NA50 Collaboration is shown schematically in Fig.~\ref{fig:setup_NA50}. The presence of additional detectors enabled the NA50 experiment to measure the collision centrality by three independent variables: $E_{T}$, $dN_{ch}/d\eta$ and $E_{ZDC}$. \\

\begin{figure}
  \includegraphics[width=1.0\linewidth]{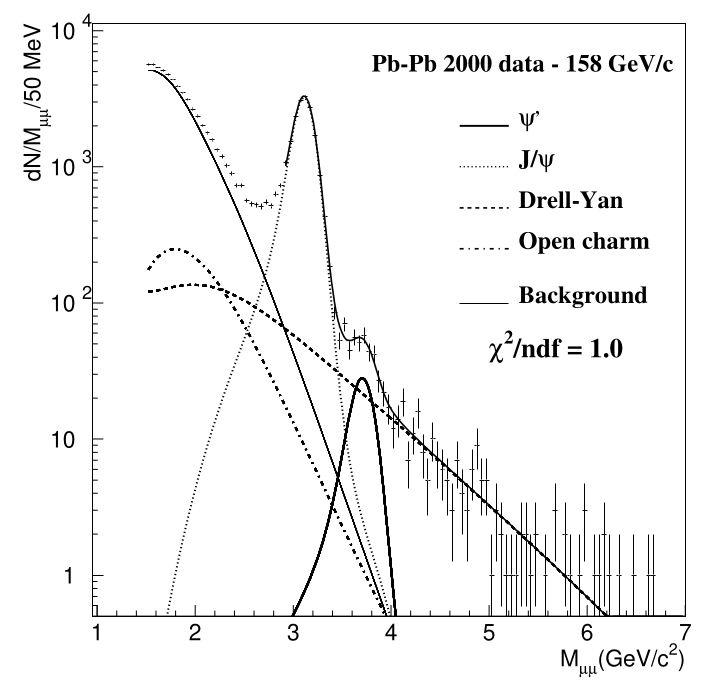}
\caption{The raw invariant mass spectrum of opposite-sign di-muons ($\mu^{+}\mu^{-}$) for 158 A GeV Pb+Pb semi-central ($35 < E_{T} < 45$ GeV) collisions measured by the NA50 experiment and reconstructed from the data sample collected in the year 2000~\cite{NA50:2004sgj}. The lines denote the fit to different sources (see text for details).}
\label{fig:invmass_spectra}       
\end{figure}

The first systematic measurement of charmonium ($J/\psi$ and $\psi(2S)$) production at CERN SPS was carried out by the NA38~\cite{NA38:1998lyg,NA38:1998udo} experiment in reactions induced by proton and light ion (O, S) beams. The p+A data were collected with proton beams of energy 450 and 200 GeV, whereas the ion beams were accelerated to a beam energy of 200 A GeV. With liquid hydrogen and deuterium targets, charmonium production was also measured in 450 GeV p+p and p+d reactions, in the frame of the NA51~\cite{NA51:1998uun} experiment. 
The measured $J/\psi$ production cross-section per nucleon-nucleon collision was seen to exhibit a remarkably continuously falling trend, when plotted as a function of the product of the mass numbers of the colliding nuclei, from p+p up to S+U interactions~\cite{Abreu:1999nn}. A power law parametrization of the production cross section, $\sigma_{AB} = \sigma_{0} (A \times B)^{\alpha}$, where $A$ and $B$ respectively denote the mass numbers of projectile and target nuclei, revealed $\alpha^{J/\psi} = 0.918 \pm 0.015$ from the global fit of p+A and A+B data, as shown in Fig.~\ref{fig:jpsi_na38}. Within S+U collisions, a similar pattern was found as a function of the collision centrality. Due to its much smaller production cross section as compared to $J/\psi$, $\psi(2S)$ measurements were affected by large statistical uncertainties ($\sim 10 \%$). To minimize the systematic uncertainties, the $\psi(2S)/J/\psi$ production cross section ratios were analyzed. $\psi(2S)$ was seen to have similar target mass dependence as $J/\psi$ in p+A collisions, but in S+U collisions the $\psi(2S)$ yield showed a clearly visible departure from this common behavior, and was significantly smaller than what was expected from the extrapolation of the p+A results suggested~\cite{NA50:2006yzz}. \\

\begin{figure*}
  \includegraphics[width=1.0\linewidth]{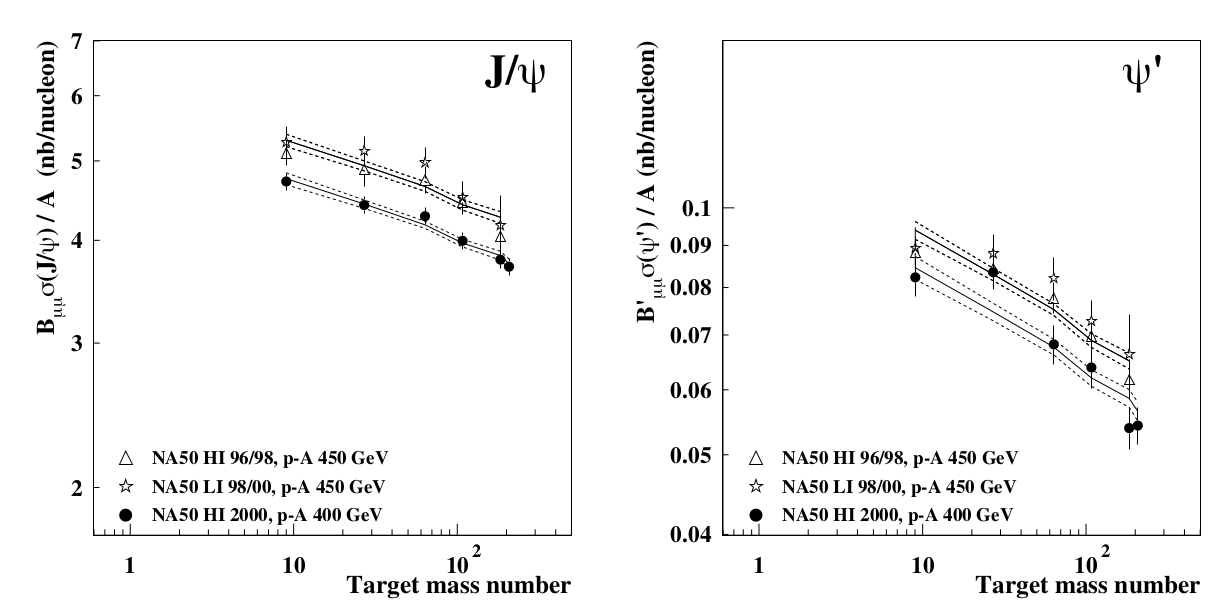}
\caption{$J/\psi$ (left) and $\psi(2S)$ (right) production cross-sections per target nucleon measured in the di-muon ($\mu^{+}\mu^{-}$) decay channel, from the three different NA50 data sets in 400 GeV and 450 GeV p+A collisions analyzed in~\cite{NA50:2006rdp}. The continuous line denotes the result of a simultaneous fit using Glauber model fit with  two independent parameters for normalisation and one common parameter for absorption cross-section. The error bands represented by dotted lines, reflect the fit parameter uncertainties associated with absorption cross-section and normalisation.}
\label{fig:na50_pA}       
\end{figure*}

Subsequently, the NA50 experiment extensively studied $J/\psi$ and $\psi(2S)$ production in p+A and Pb+Pb collisions. Data were published in terms of $J/\psi$-to-Drell-Yan (DY) ratio of cross sections, with muons from DY process acting as a reference. Being electromagnetic in nature, the DY di-muons are not affected by the strong interaction and thus considered as a robust reference for normalization while studying centrality dependent $J/\psi$ production in nuclear collisions. Being collected under identical experimental conditions as of $J/\psi$, the ratio helps to minimize the systematic uncertainties in the data by canceling the uncertainties originating from various sources like fluctuations in beam luminosity, experimental bias, detector inefficiency and reconstruction criteria. The only disadvantage of using this ratio is the much smaller DY cross section as compared to $J/\psi$ resulting in significantly larger statistical uncertainties associated to the ratio than to the absolute $J/\psi$ yield. Fig.~\ref{fig:invmass_spectra} shows a typical oppositely charged di-muon invariant mass spectrum reconstructed from the NA50 data sample collected in the year 2000, for Pb+Pb collisions. $J/\psi$, $\psi(2S)$ and DY yields were obtained from a fit to the mass spectrum according to the formula:

\begin{eqnarray}
 &{dN^{+-} \over dM_{\mu\mu}}   =  A_{J/\psi}{dN^{J/\psi} \over dM_{\mu\mu}} + A_{\psi(2S)}{dN^{\psi(2S)} \over dM_{\mu\mu}} + \nonumber \\ & A_{DY}{dN^{DY} \over dM_{\mu\mu}} + A_{D\bar{D}}{dN^{D\bar{D}} \over dM_{\mu\mu}} + {dN^{BG} \over dM_{\mu\mu}}    
\end{eqnarray}
\par
by accounting for different sources contributing to the overall di-muon spectrum. The sources included the di-muon decay of $J/\psi$ (BR $\sim  5.94 \times 10^{-2}$) and $\psi(2S)$ (BR $\sim 7.3 \times 10^{-3}$) resonances as well as continuum from the DY process,  the semi-leptonic decay of correlated open charm hadrons ($D\bar{D}$) and the nonphysical combinatorial background mostly originating from the uncorrelated weak decay of charged pions and kaons. Dedicated Monte Carlo (MC) simulations were performed for disentangling 
the various contributions and to provide the required ingredients to determine cross-sections. Such techniques were found useful to evaluate the detector acceptance ($A$) of each physical process and the momentum resolution of the associated muon pair candidates as they resulted from the real detector set up and selection criteria applied on the data. Di-muons from each source were generated separately and transported through the simulated experimental configuration similar to the real muons, accounting in particular for the energy loss and multiple scattering inside the hadron absorbers and for the limited geometrical coverage of the spectrometer. The resulting smearing of the reconstructed muon momentum, due to the experimental resolution of the apparatus, was directly reflected in the shape of the invariant mass distributions of the reconstructed $J/\psi$ and $\psi(2S)$  mesons, which were modeled using Gaussian functions with asymmetric tails with an approximate mass resolution of $100$ MeV/$c^2$.
On the other hand, the shape and the normalization of the combinatorial background
was entirely determined from the like-sign pair distributions in the measured data sample, according to, $N_{BG}^{\pm} = 2\sqrt{N^{++}N^{--}}$. The use of the like-sign method requires the spectrometer acceptance to be independent of the charge of the reconstructed muon tracks, which was ensured by the applied event selections. The signal yields are extracted 
by performing a fit to the mass spectrum in the range $2.9 < M_{\mu\mu} < 4.5$ GeV/$c^2$, where the charmonium yields are least sensitive to the other sources present in the spectrum. With this "standard" analysis strategy NA50 measured $J/\psi$ and $\psi(2S)$ production in 400 GeV and 450 GeV p+A and 158 A GeV Pb+Pb collisions. 
 
  With 450 GeV proton beams, NA50 measured $J/\psi$ and $\psi(2S)$ production with five different (Be, Al, Cu, Ag, W) nuclear targets~\cite{NA50:2003pvd,NA50:2003fvu}. For each target two statistically independent data samples were collected. They are referred to as high intensity (HI) and low intensity (LI) sets, corresponding to average beam fluxes of $2 - 3 \times 10^{9}$ and $3 - 6 \times 10^{8}$ protons during 2.37 s bursts, respectively. The HI data set was found to be more suitable to study the $\psi(2S)$ state and the shapes of differential distributions. To enable a coherent comparison both the data samples at 450 GeV are evaluated within an identical kinematic window: $-0.5 < y_{cms} <0.5$ and $-0.5 < cos(\theta_{cs}) < 0.5$, where $\theta_{cs}$ is the Collins-Soper angle. It denotes the polar angle of the $\mu^{+}$ relative to the bisector of the angle between the projectile and target momentum directions in the di-muon rest frame. NA50 also measured $J/\psi$, $\psi(2S)$ and DY production in the kinematical domain, $-0.425 < y_{cms} < 0.575 $ and $-0.5 < cos(\theta_{cs}) < 0.5$ using 400 GeV proton beams incident on Be, Al, Cu, Ag, W and Pb targets~\cite{NA50:2006rdp}. The intensity of 400 GeV proton beams were even slightly higher than that of 450 GeV HI data samples. However the availability of limited beam time period prevented the recording of sizable statistics. However, unlike previous measurements, during the 400 GeV run, the data for the different nuclear targets were collected in the same data-taking period,  which significantly reduced the systematic uncertainties. \\

\begin{figure*}
  \includegraphics[width=1.0\linewidth]{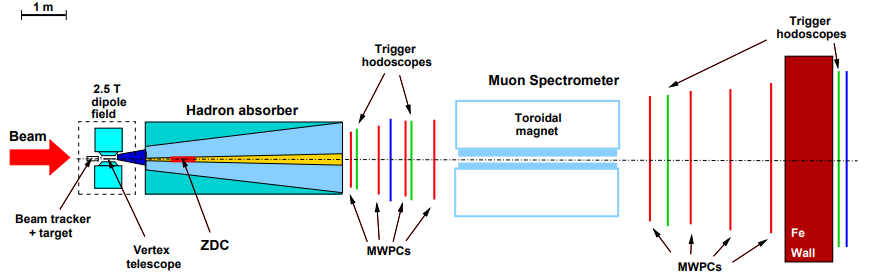}
\caption{Schematic of the muon spectrometer employed by {\bf the} NA60 experiment at CERN SPS facility, for dimuon measurements in 400 GeV and 158 GeV p+A and 158 A GeV In+In collisions~\cite{Scomparin:2007rt}.}
\label{fig:setup_NA60}       
\end{figure*}
 
  One common feature observed at both the energies was that the production cross sections of $J/\psi$ and $\psi(2S)$ states in p+A collisions decreased with increasing size of the target nucleus. The yield of DY di-muons on the other hand had been seen to exhibit a linear scaling with the target mass number. Within the framework of $\alpha$ parameterization, a combined fit of the mass dependence of the charmonium production cross sections with HI and and LI data sets at 450 GeV resulted in $\alpha_{J/\psi}^{450} = 0.928 \pm 0.015$ and $\alpha_{\psi(2S)}^{450} = 0.888 \pm 0.018$. The corresponding values obtained for 400 GeV p+A collisions were $\alpha_{J/\psi}^{400} = 0.925 \pm 0.009$ and $\alpha_{\psi(2S)}^{400} = 0.852 \pm 0.019$. The results indicate stronger suppression of $\psi(2S)$ states in the nuclear medium. Although widely used in experimental analyses, the $\alpha$ parametrization represents a rough estimate of CNM effects, as the extracted value of $\alpha$ was found to be sensitive to the nucleus used as the lightest target. \\

The Glauber model~\cite{Bialas:1977pd,Glauber:1970jm,Czyz:1969jg,Bialas:1976ed} offers a more rigorous formalism for describing the observed charmonium production cross-sections. Within this formalism, a p+A or A+A collision is considered to be an incoherent sum of independent interactions between the nucleons from the projectile and the target. Nucleon properties are assumed to be unchanged after the first collision such that they undergo the next nucleon-nucleon interactions with the same cross section. The charmonium production cross section per nucleon in p+A collisions can be expressed as:

\begin{equation}
\sigma_{pA}={\sigma_{0} \over \sigma_{abs}^{G}} \int{d\vec{b} \left[1 - \left( 1 - T_{A}\sigma_{abs}^{G} \right)^{A}\right]}
\label{Eq:2}
\end{equation}

where $\sigma_{0}$ denotes normalization and the parameter $\sigma_{abs}^{G}$ represents that absorption cross section of the initially produced $c\bar{c}$ pairs in their pre-resonant or resonance state while they traverse through the nuclear matter. $T_{A}(\vec{b})$ denotes the nuclear thickness parameter at an impact parameter $\vec{b}$. The thickness parameter $T_{A}(\vec{b})$ represents nuclear density per unit surface area and can be estimated from the nuclear density profile. Often the detailed Glauber model formalism is approximated, in first order, by the so called "$\rho L$ parametrization" as 

\begin{equation}
\sigma_{pA} = \sigma_{0}^{\rho L}  A  exp(-\sigma_{abs}^{\rho L}<\rho L>)
\label{Eq:3}
\end{equation}

where $\sigma_{abs}^{\rho L}$ represents the (approximate) absorption cross-section of the charmonium states, $\rho$ denotes the nuclear matter density and $L$ is the mean distance traveled by the resonance states on their way out of the target nucleus and calculable using the Glauber model for each target nucleus. At 450 GeV, a combined fit to LI and HI data sets based on the $\rho L$ parametrization resulted in absorption cross sections of $\sigma_{abs}^{\rho L} = 4.3 \pm 0.7$ mb for $J/\psi$ and $\sigma_{abs}^{\rho L} = 6.6 \pm 0.8$ mb for $\psi(2S)$ states~\cite{NA50:2003fvu}. The corresponding analysis at 400 GeV reported an absorption cross section of $4.2 \pm 0.5$ ($4.6 \pm 0.6$) mb for $J/\psi$ and $8.3 \pm 1.0$ ( $10. 1 \pm 1.6$) mb for $\psi(2S)$ charmonium states using $\rho L$ parametrization (full Glauber model). A combined fit to all three p+A data sets at 400 GeV and 450 GeV, following full Glauber model, yielded $\sigma_{abs}^{G}(J/\psi) = 4.5 \pm 0.5$ mb and $\sigma_{abs}^{G} (\psi(2S)) = 8.3 \pm 0.9$ mb, as illustrated in Fig.~\ref{fig:na50_pA}~\cite{NA50:2006rdp}. The results clearly showed the stronger nuclear absorption suffered by $\psi(2S)$ than $J/\psi$. The magnitude of the absorption cross section is also found to agree well with the values independently extracted from $\psi$-to-DY ratios confirming the robustness of the results. \\ 

\begin{figure}
  \includegraphics[width=1.0\linewidth] {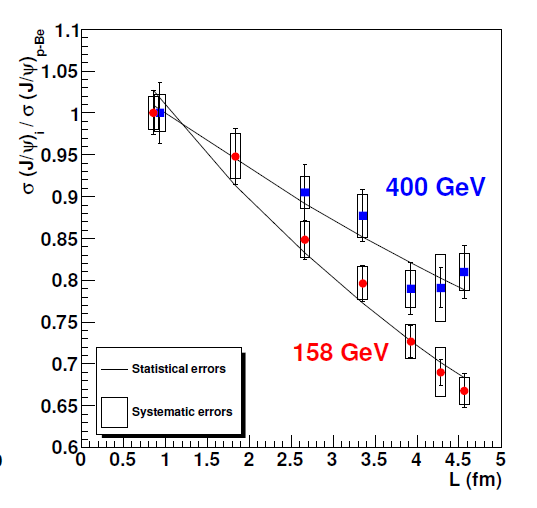}
\caption{Ratio of the $J/\psi$ production cross sections in p+A collisions (for 7 different target nuclei) and p+Be collisions at 158 GeV (circles) and 400 GeV (squares) as a function of $L$, measured by the NA60 experiment at CERN-SPS facility~\cite{Scomparin:2009tg,Arnaldi:2009ph}. The systematic errors represented by boxes include contributions from uncertainties on target thicknesses, di-muon reconstruction efficiency and on the shape of the rapidity ($y_{cm}$) distribution employed for acceptance correction. Only the fraction of the total systematic error uncommon to all the data points, which affects the evaluation of nuclear effects, is quoted. The lines represent the Glauber model fit to the data.}
\label{fig:na60_pA}       
\end{figure}

The most recent measurements of charmonium production in p+A collisions at SPS have been carried out by the NA60 experiment~\cite{Scomparin:2007rt,Scomparin:2010zz,Scomparin:2009tg}. $J/\psi$ production in p+A collisions was recorded at 158 GeV beam energy for the first time. The main objective was to build a reference {\bf in the same} energy and kinematic conditions as heavy-ion data. As a second generation fixed target experiment, NA60 strongly augmented the di-muon detection techniques as compared to its predecessors. The essential components of the NA60 experimental setup are shown in Fig.~\ref{fig:setup_NA60}. The muon spectrometer and the ZDC were adopted from the NA50 experiment. However, the vertex region was thoroughly upgraded, by inclusion of a silicon pixel layer based vertex telescope (VT), placed within a dipole magnetic field of strength 2.5 T and a beam tracker (BT), to measure the transverse position of the incident beam particles before interacting with the target. This redesign of the vertex region led to the improved capabilities to the determination of the primary vertex with a spatial resolution 10–15 $\mu$m in the transverse plane and $\sim  200$ $\mu$m along the longitudinal axis. The tracks reconstructed in the two spectrometers were simultaneously matched both in coordinate and momentum space to obtain an accurate measurement of the muon kinematics and to minimize the uncertainties induced by energy loss fluctuations and multiple scattering due to the passage of the muon candidate tracks through the hadron absorber, as experienced by previous experiments. This ultimately led to a better pair mass resolution of $\sim  70$ MeV in the $J/\psi$ mass region. An improved vertex resolution also helped to isolate the prompt di-muons from the inclusive sample. During the p+A data taking period, the target system of NA60, was comprised of 7 different target nuclei (Be, Al, Cu, In, W, Pb, U) simultaneously exposed to the incident beam of intensity $\sim 5 \times 10^{8}$ protons/s. The experiment also recorded data at 400 GeV, using the identical detector configuration as the one used in the 158 GeV data taking period, in order to  cross-check the NA50 results obtained at the same energy. \\

\begin{figure}
  \includegraphics[width=1.0\linewidth] {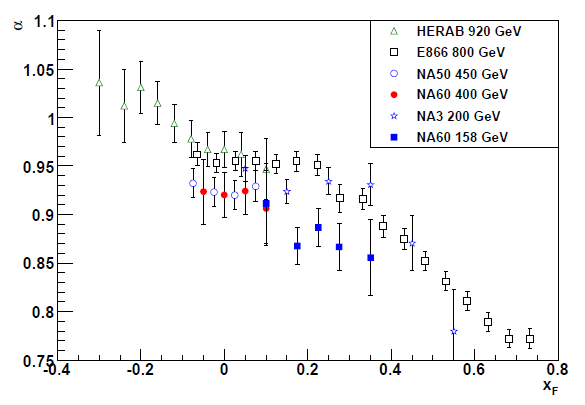}
\caption{Compilation of the values of the $\alpha$-parameter as a function of $x_F$~\cite{Scomparin:2009tg,Arnaldi:2009ph}. The $\alpha$ values are extracted from the power law fit ($A^{\alpha}$) to the target mass ($A$) dependence of the $J/\psi$ production cross sections in p+A collisions as measured by various fixed target collision experiments at different energies.}
\label{fig:alpha_xF}       
\end{figure}

\begin{figure*}
  \includegraphics[width=0.5\linewidth]{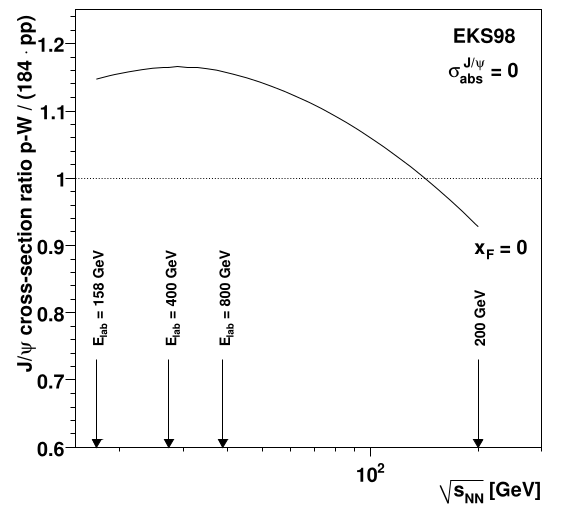}
     \includegraphics[width=0.5\linewidth]{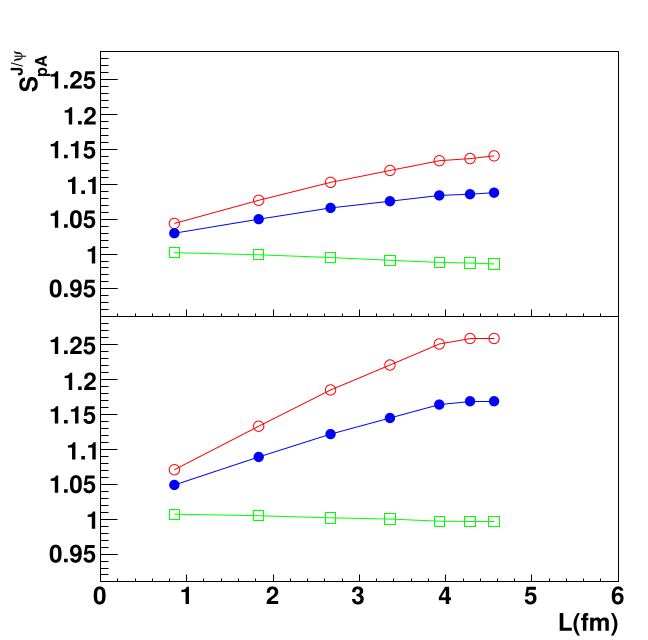}
\caption{Ratio of $J/\psi$ production cross section at mid-rapidity ($x_F =0$) in p+W and pp collisions calculated with EKS98 nPDF scheme, scaled to the Tungsten mass number, and shown as a function of the nucleon-nucleon centre-of-mass energy ($\sqrt{s_{NN}}$). The figure is adopted  from~\cite{Lourenco:2008sk}. (right) Variation of the central rapidity $J/\psi$ cross section due to nuclear PDFs with the size of nuclear target in 158 GeV p+A collisions. Open squares and circles respectively denote contributions of $q\bar{q}$ annihilation and $gg$ fusion processes to the total cross section denoted by filled circles and calculated with the EKS98 (top panel) and EPSS08 (bottom panel) nPDF schemes. The figure is adopted from~\cite{Arnaldi:2009it}.}
\label{fig:shadowing_lourenco}       
\end{figure*}

The analysis of the $J/\psi$ production cross sections at 158 GeV was performed in the rapidity range $0.28 < y_{cms} < 0.78$. Due to the large statistical uncertainties associated with the DY pairs, results were published in terms of cross section ratios $\sigma^{J/\psi}_{pA}/\sigma^{J/\psi}_{pBe}$ between the target of mass number A and the lightest one (Be). The ratio was useful to get a better control over the systematics by largely canceling the beam luminosity factors. For minimizing the relative systematic errors, data at 400 GeV were recorded with the identical experimental configuration and analyzed in the rapidity range $-0.17 < y_{cms} < 0.33 $, which corresponded to the same laboratory frame rapidity of the 158 GeV data. The corresponding results are displayed in Fig.~\ref{fig:na60_pA} as a function of $L$. The target mass dependence of the 158 GeV data sample is clearly seen to be steeper than the 400 GeV p+A data. The systematic errors associated with each data point include contributions from uncertainties on target thicknesses, on the shape of the rapidity distribution used as input in the acceptance estimation and on the efficiency of the reconstructed single muon tracks. Analyzing these data using the $\alpha$ parametrization produced $\alpha_{J/\psi} =0.882 \pm 0.009 \pm 0.008$ at 158 GeV and $\alpha_{J/\psi} = 0.927 \pm 0.13 \pm 0.009 $ at 400 GeV, indicating larger suppression at lower incident proton beam energy. Alternatively, a Glauber model analysis of the target dependence of the cross-section ratios generates $\sigma_{abs}^{\rho L} = 7.6 \pm 0.7 \pm 0.6$ mb at 158 GeV and $\sigma_{abs}^{\rho L} = 4.3 \pm 0.8 \pm 0.6$ mb at 400 GeV, showing stronger suppression at lower incident proton beam energies~\cite{Scomparin:2009tg,Arnaldi:2009ph}. The value of $\sigma_{abs}^{\rho L}$ at 400 GeV extracted by the NA60 experiment was found to be in excellent agreement with the corresponding value reported by the NA50 experiment. The observed dependence of the nuclear effects in p+A collisions on the incident proton beam energy was further investigated by comparing the NA60 results with previous measurements performed at fixed target collisions.
For this purpose, a compilation of the measured values of the $\alpha$ parameter was performed as a function of $x_{F}$. The results are shown in Fig.~\ref{fig:alpha_xF} and they include measurements from NA50 at 450 GeV, E866 at 800 GeV and HERA-B at 920 GeV. At a fixed energy, smaller $\alpha$ at larger $x_{F}$ indicates stronger suppression effects. At a given $x_{F}$, an enhanced suppression occurs as the proton beam energy is lowered. \\
\begin{figure*}
  \includegraphics[width=0.5\linewidth]{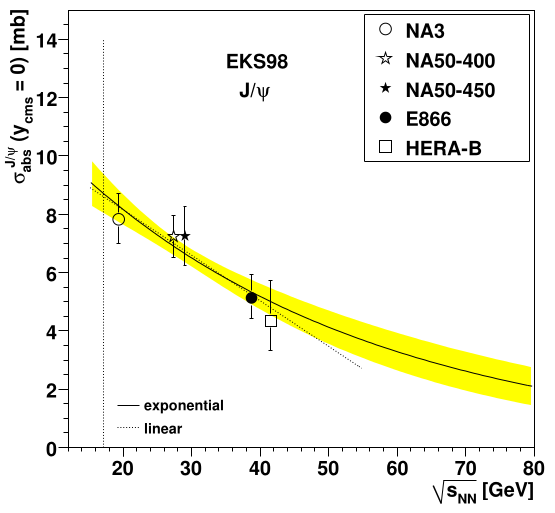}
  \includegraphics[width=0.5\linewidth]{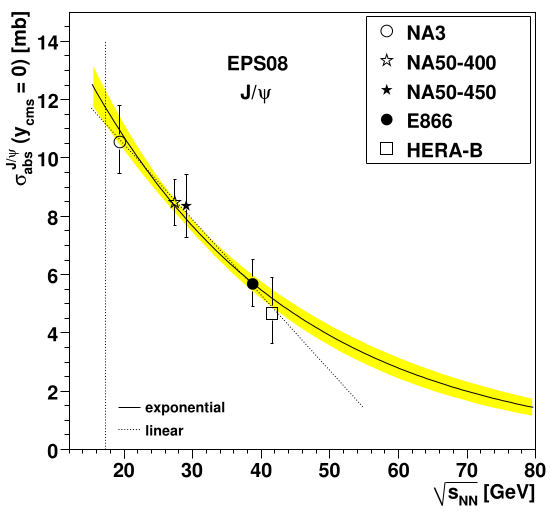}
\caption{$\sqrt{s_{NN}}$ dependence of the mid-rapidity ($y_{cms} =0$) final state $J/\psi$ absorption cross section, as reported in~\cite{Lourenco:2008sk}. $\sigma_{abs}^{J/\psi}$ values are extracted by fitting mid-rapidity $J/\psi$ production cross sections in fixed target p+A collisions at different energies within a Glauber model framework using EKS98 (left) and EPS08 (right) nPDF schemes to model the nuclear modification of parton densities inside the target nuclei.}
\label{fig:sigma_abs_Eb}       
\end{figure*}

It is worthwhile to note here that both $\alpha_{J/\psi}$ and $\sigma_{abs}^{\rho L}$ directly extracted from the data, are essentially effective quantities\footnote{Hence, now onward, we denote the nuclear absorption cross section, directly extracted from data, without incorporating any other CNM effect, by $\sigma_{abs}^{eff}$.}. They characterize the comprehensive influence of various cold nuclear matter effects present in the data due to a complex interplay of nuclear modification of parton densities, initial state energy loss and final stage absorption of the initially produced $c\bar{c}$ pairs inside the target nucleus. Detailed phenomenological modelings as attempted in~\cite{Vogt:1999dw,Lourenco:2008sk,Arnaldi:2009it,Bhaduri:2014daa,Giri:2026tps} are required to isolate the role of each of these effects to describe the overall suppression pattern observed in the data. The then available $J/\psi$ production cross sections measured in fixed target p+A collisions, in the proton beam energy range between 200 to 920 GeV, and in d+Au collisions at RHIC, at $\sqrt{s_{NN}} = 200$ GeV have been analyzed in~\cite{Lourenco:2008sk} within the Glauber model framework employing different sets of parton densities with and without nuclear modifications. Analysis results exhibit a strong sensitivity of the final state absorption cross section on the $J/\psi$ kinematics and on the collision energy. At energies relevant to fixed target experiments, charmonium production probes Bjorken-$x$ values in the “anti-shadowing” region characterized by enhanced parton densities inside the heavy nucleus. This leads to an enhanced $c\bar{c}$ cross section at the initial stage of $J/\psi$ production in p+A collisions as compared to collisions between free protons. This is displayed in the left panel of Fig.~\ref{fig:shadowing_lourenco}, where the so called shadowing factor defined as the ratio of per nucleon $J/\psi$ ($c\bar{c}$) production cross sections in absence of any final state absorption in p+A collision to that in p+p collision, is plotted at mid-rapidity ($x_F =0$) as a function of collision energy, for p+W collisions using the EKS98 nuclear parton distribution function (nPDF). While the presence of an anti-shadowing effect in the kinematic region probed by charmonium measurements at SPS energy domain is true for almost all standard nPDF schemes, the degree of enhancement depends on the employed nPDF model and the effect becomes stronger with the size of the target nucleus. This is illustrated in the right panel of Fig.~\ref{fig:shadowing_lourenco}, where the mid-rapidity shadowing factor is plotted as a function of $L$ for different nuclear targets from Be to U, in 158 GeV p+A collisions using the EKS98 and EPS08 nPDF schemes. Shadowing factors for the partonic sub-processes $gg$ fusion and $q\bar{q}$ annihilation are separately shown in addition to their total contribution. Interestingly 
for both the nPDF sets, the cross section due to $q\bar{q}$ annihilation exhibits a small shadowing, which is more than counterbalanced by the dominant $gg$ fusion process. Also the net anti-shadowing effect appears to be more pronounced with the EPS08 model. To balance this enhanced $c\bar{c}$ production, a larger value of final state absorption cross section ($\sigma_{abs}^{J/\psi}$), as compared to corresponding $\sigma_{abs}^{eff}$ obtained from a pure Glauber model analysis, is required to reproduce the suppression pattern present in the data\footnote{$\sigma_{abs}^{eff}$ is a phenomenological parameter which accounts for the interplay of the different CNM effects (nPDF, energy loss, nuclear absorption) and obtained from a direct fit to the data. $\sigma_{abs}^{J/\psi}$ also obtained from phenomenological modeling, represents the dissociation probability of $J/\psi$ mesons (or the pre-resonant $c\bar{c}$ pairs) by nucleons, once the other CNM effects (nPDF in particular) are explicitly incorporated in the analysis.}. At a fixed collision energy, $\sigma_{abs}^{J/\psi}$ shows {\bf a} significant dependence on the $J/\psi$ rapidity, 
once the nuclear modification of parton densities in the initial stage are taken into account~\cite{Lourenco:2008sk}. The level of final state absorption of mid-rapidity $J/\psi$ has been found to increase with decreasing collision energy. However the specific numerical values of $\sigma_{abs}^{J/\psi}$ are found to be sensitive to the employed nPDF set, with larger absorption cross sections for EPS08 model as compared to EKS98 model, as illustrated in Fig.~\ref{fig:sigma_abs_Eb}. But the decrease of absorption cross section with increasing energy of collision is found to be a general feature, subsequently verified experimentally, once the NA60 data on 158 GeV p+A collisions have become available. It would be worthwhile to recall here the main lessons learnt from the charmonium measurements in p+A collisions at SPS energies. A significant suppression of charmonium production is observed in p+A collisions with more suppression for heavier target nuclei. The suppression is usually attributed to the convolution of the various CNM effects including nuclear modification of the target parton densities in the initial stage and final state dissociation of the initially produced $c\bar{c}$ pairs due to interactions with the projectile and the target nucleons or with the particles produced in the collision. The measured suppression in production cross section is commonly quantified through an "effective" nuclear absorption cross section which has a non-negligible kinematic dependence and grows stronger with decreasing energy of the collision. Therefore, in order to use it as a baseline for A+A collisions, the p+A reference should be built from the measurements in the same energy and kinematic domain of heavy-ion collisions. \\

\begin{figure*}
  \includegraphics[width=0.5 \linewidth]{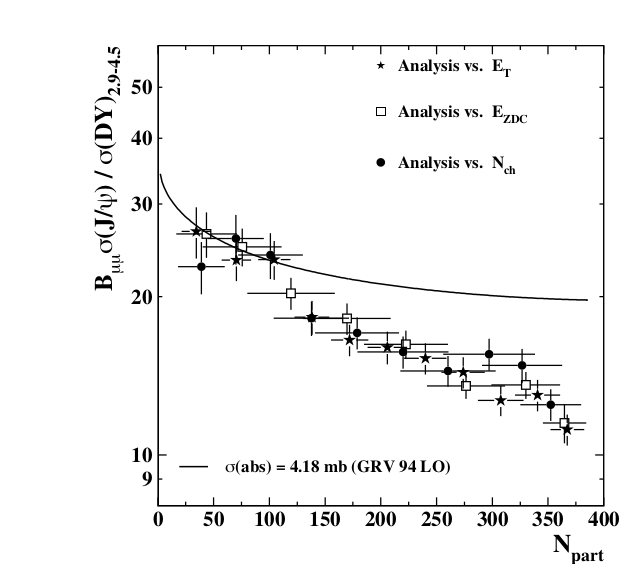}
  \includegraphics[width=0.45 \linewidth]{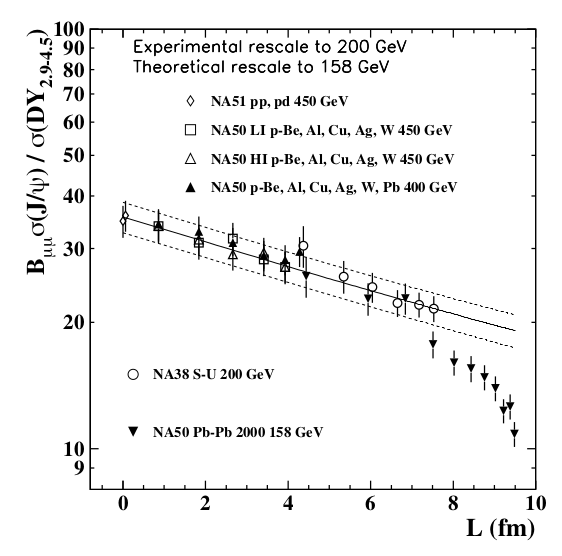}
\caption{(left) $J/\psi$/Drell-Yan (DY) cross-section ratios as a function of $N_{part}$, independently analyzed using three centrality estimators ($E_{T}$, $E_{ZDC}$ and $N_{ch}$) measured by the NA50 experiment in 158 A GeV Pb+Pb collisions at CERN SPS facility. The line corresponds to the normal nuclear absorption pattern evaluated using pure Glauber model with an input absorption cross section of $\sim 4.2$ mb. (right) The $J/\psi$/Drell-Yan (DY) cross-section ratio as a function of $L$, for several collision systems, measured by the NA50 experiment. The result of the analysis as a function of $E_{T}$ was used for the Pb+Pb data points. For p+A collisions, a rescaling of the measured data has been performed to 158 GeV/nucleon. Data are compared to the normal nuclear absorption pattern. The band corresponds to the uncertainties associated with the parameters of the Glauber model. The figures are selected from~\cite{NA50:2004sgj}}.
\label{fig:jpsi_na50_PbPb}       
\end{figure*}

After discussing charmonium production in p+A interactions we now move on to discuss the corresponding measurements in heavy-ion collisions. The ultimate goal of measuring charmonium production in A+A collisions is to look for suppression effects that can be connected to the formation of a deconfined medium in such collisions. However as discussed above, the $J/\psi$ (and also the $\psi(2S)$) yield already suffers considerable depletion in p+A collisions owing to the CNM effects. This nuclear suppression also remains operative in case of heavy
ion collisions, even at a higher degree due to the amplified cold matter effects induced by both
target as well as projectile nucleons. If some “anomalous’ suppression in addition to the normal nuclear suppression can be detected in the heavy-ion data, then that can be connected to the presence of a medium eventually produced in the collisions. A charmonium resonance created in such collisions will first interact with the nucleons of the colliding ions, as in p+A collisions. Then, after the nuclei have separated, the charmonium will interact with the medium produced in the collision. \\ 

\begin{figure*}
  \includegraphics[width=0.3 \linewidth]{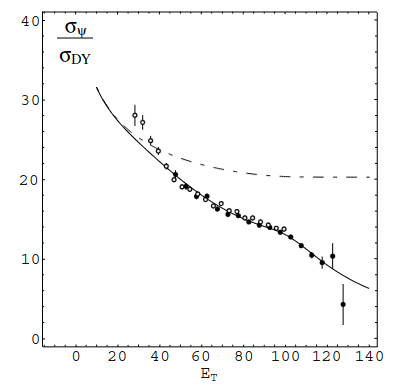}
  \includegraphics[width=0.3 \linewidth]{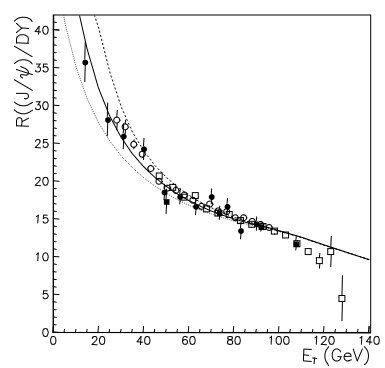}
  \includegraphics[width=0.3 \linewidth]{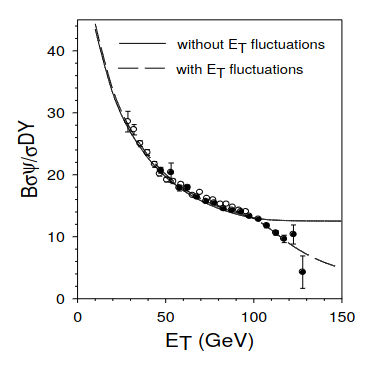}
\caption{(left) $J/\psi$ suppression in Pb+Pb collisions at SPS energy. The solid line denotes the calculation within the static threshold model scenario and the dashed line denotes result only with nuclear absorption. The figure is taken from~\cite{bl00}. (middle) $J/\psi$ suppression in SPS Pb+Pb collisions within the hadronic comover absorption scenario. The figure is adopted from~\cite{ca00}. It is important to note that, both the models assume a nuclear absorption cross section of about 4.2 mb based on the p+A collisions measured by the NA50 experiment at incident proton beam energies of 400 and 450 GeV. Thus no firm conclusion can be drawn from these models in the light of the subsequent determination of nuclear absorption cross section in 158 GeV p+A collisions by the NA60 experiment. (right) $J/\psi$ suppression in 158 A GeV Pb+Pb collisions within the QVZ model scenario. Solid and dashed lines respectively denote model calculations without and with $E_{T}$ fluctuations. The figure is taken from~\cite{ch02b}.}  
\label{fig:jpsi_sps_model}       
\end{figure*}

The first significant
measurement of $J/\psi$ suppression in heavy-ion collisions was performed by the NA50 Collaboration in 158 A GeV Pb+Pb collisions, in the di-muon kinematic domain, $0 < y_{cms} < 1$. The data were collected in different runs during the period 1995 - 2000. 
The latest data sample for Pb+Pb interactions was collected by the NA50 experiment in the year 2000, under improved experimental conditions with respect to previous runs. 
With use of one single Pb target placed in vacuum the setup was better suited to efficiently solve the issues affecting previous data collections like re-interaction of the projectile fragments inside the target assembly leading to a centrality smearing as well as Pb-air interactions contaminating the peripheral Pb+Pb collisions~\cite{NA50:2000brc}. 
Data were recorded with an incoming beam intensity of $(1 - 1.4) \times 10^{7}$ ions/s over 4.8 s bursts with 20 s interval, incident on a 4 mm thick  target, corresponding to an interaction probability of $\sim 10 \%$. 
Muon pairs were selected in the phase space domain $2.92 < y < 3.92$ ($0 \le y_{cms} < 1$) and $|cos\theta_{CS}|< 0.5$,
resulting in an acceptance of $\sim 14 \%$.  Analysis of the data sample showed that the $J/\psi$/DY cross section ratio ($B_{\mu\mu}\sigma_{J/\psi}/\sigma_{DY}$) measured in the most peripheral Pb+Pb interactions was seen to be in good agreement with the pattern of normal nuclear absorption, with $\sigma_{abs}^{eff} \simeq 4.2$ mb, as deduced using pure Glauber model framework, from p+A collision data alone measured by the same experiment with incident proton beams at the energies of 400 and 450 GeV. 
For semi-central Pb+Pb collisions the ratio departed from the normal nuclear absorption pattern and continuously decreased with increasing centrality generating $30 - 40 \%$ additional suppression in the most central collisions. Results obtained from three independent analyses, using the centrality variables $E_{T}$, $E_{ZDC}$ and $N_{ch}$ were found to be very consistent, with onset of anomalous suppression beyond $N_{part} \sim 100$, as presented in the left panel of Fig.~\ref{fig:jpsi_na50_PbPb}, using $N_{part}$, the average number of participant nucleons, as the common centrality variable. A comparison of the Pb+Pb results with other lighter collision systems is shown in the right panel of Fig.~\ref{fig:jpsi_na50_PbPb}, the $J/\psi$/DY ratio is plotted as a function of $L$. An anomalous $J/\psi$ suppression with respect to the reference curve obtained from p+A data is clearly visible for semicentral and central Pb+Pb collisions. Note that the reference curve was obtained using a nuclear absorption cross-section $\sigma_{abs}^{J/\psi} \sim  4.2$ mb and assuming the nuclear absorption to be independent of beam energy and $J/\psi$ rapidity.\footnote{These analyses were performed prior to the NA60 measurement of $J/\psi$ production in 158 GeV p+A collisions.} The S+U data points from the NA38 experiment closely matched the nuclear absorption curve, leaving hardly any room for anomalous suppression in S+U collisions. \\

\begin{figure*}
  \includegraphics[width=0.49 \linewidth]{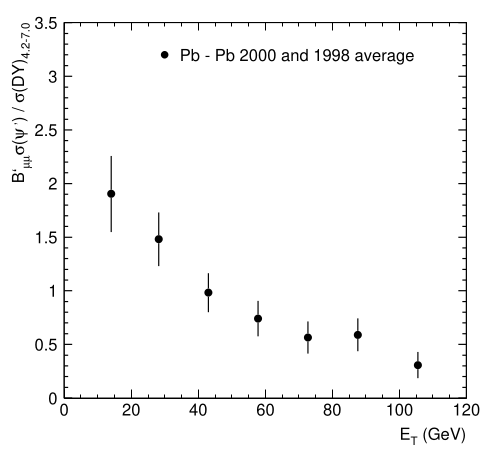}
  \includegraphics[width=0.49 \linewidth]{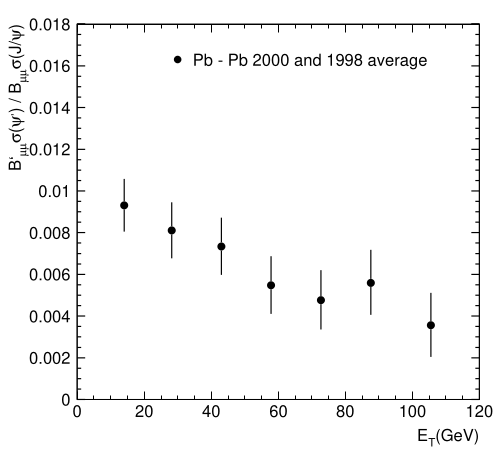}
\caption{(left) $\psi(2S)$ (designated by $\psi'$ in the axis labels)/Drell-Yan cross section ratio as a function of collision centrality  (measured in terms of $E_{T}$) in 158 A GeV Pb+Pb collisions, for combined data samples collected in the years 1998 and 2000. (right) Centrality dependence of $\psi(2S)$/$J/\psi$ cross section ratio for the same set of data samples. Both the results are adopted from~\cite{NA50:2006yzz}.}
\label{fig:psi_prime_na50_PbPb}       
\end{figure*}

Such anomalous suppression of $J/\psi$ yield beyond that expected from CNM effects was expected to carry the evidence for de-confinement transition at SPS. However the physical origin of this suppression is till date a debated issue, as different models with and without considering the presence of a de-confined medium were found to explain the data. In the geometrical threshold model~\cite{bl96,bl00} developed in the spirit of Debye screening phenomena, without implementing the microscopic dynamics, the $J/\psi$ survival probability in nuclear collisions at an impact parameter $b$, is modeled as:

\begin{equation}
S_{J/\psi}(b) = \int{d^2\vec{s} S_{J/\psi}^{nucl}(b,\vec{s}) \Theta(n_{c} - n_{part}(b,\vec{s}))}
\label{Eq:surv_TH}
\end{equation}

where $S_{J/\psi}^{nucl}(b,\vec{s})$ is the survival probability of $J/\psi$ mesons suffering nuclear absorption at a position (b,\vec{s}) in the transverse plane. The participant density $n_{part} (b,\vec{s})$ calculable from Glauber model is assumed to be proportional to the local energy density, $\epsilon(b,\vec{s})$ in the medium. In the hot and dense core of the fireball where $n_{part}$ is larger than a threshold value $n_{c}$ all the $J/\psi$ are melted, while outside in the corona region ($n_{part} < n_{c}$), the $J/\psi$ mesons suffer only nuclear dissociation. Assuming the critical density, $n_{c}$, to be equal to the maximum value of $n_{part}$ in S+U collisions, the model was found to explain the anomalous suppression in Pb+Pb collisions, see Fig.~\ref{fig:jpsi_sps_model} left. The threshold model is a static macroscopic model based on the simplified assumption of the existence of a critical threshold on the local energy density in the fireball which is proportional to the transverse density of participant nucleons. However relativistic nuclear collisions produce a finite size rapidly expanding fireball with fast decreasing temperature and short lifetime. More realistic descriptions of $J/\psi$ suppression in a deconfined medium thus include dynamical interactions between energetic partons and charmonia, that lead to sizable inelastic collision rates comparable to the cooling rate of the fireball. Within this picture charmonia can be destroyed by hard gluons even below $T_{d}$. Debye screening on the other hand controls the in medium binding energy and thus determines the effective phase space (and in medium decay width) of the dynamical dissociation reactions. At leading order gluon dissociation of $J/\psi$ ($g + J/\psi \rightarrow c + \bar{c}$) is calculated analogously to photo disintegration of electromagnetic bound states becomes dominant~\cite{Kharzeev:1994pz}. Close to $T_{d}$ the in medium binding energies are small and this leads to shrinkage of the phase space for gluon disintegration  ($g + J/\psi \rightarrow c +\bar{c}$) of  $J/\psi$ mesons. Instead the inelastic parton scattering, $g (q, \bar{q}) + J/\psi \rightarrow g (q, \bar{q}) + c + \bar{c}$ at the next-to-leading order (NLO) drives the partonic dissociation~\cite{Grand,Park:2007zza}.
Alternatively, the observed anomalous suppression of $J/\psi$ mesons at SPS was also found to be explained by dissociation induced by a confined hadronic medium. Within the hadronic co-mover scenario~\cite{Vogt:1988fj,Gavin:1988hs,ca00,Prorok:2008zm}, the secondary hadrons (eg: $\pi$, $\rho$, $\omega$) present in a hot and dense hadronic gas can inelastically scatter with the charmonia causing $J/\psi$ suppression. The additional $J/\psi$ suppression beyond the nuclear absorption in the hadronic 
comover scenario is usually modeled as 

\begin{equation}
S_{J/\psi}^{co}= exp(-\int{d\tau<v\sigma_{co}>\rho_{co}(\tau)})
\label{Eq:surv_TH}
\end{equation}

where $\rho_{co}(\tau)$ denotes the density of co-movers at the proper time $\tau$ at the position of $J/\psi$ and $<v\sigma_{co}>$ denotes the velocity dependent inelastic cross section averaged over the various co-moving hadrons and interaction energy present in the medium. Co-moving density is usually estimated from Bjorken hydrodynamical model and thus varies as $1/\tau$. The inelastic cross section $\sigma_{co}$ is an adjustable parameter in the model. With $\sigma_{co} \sim 1$ mb, the comover model was seen to describe the anomalous $J/\psi$ suppression in Pb+Pb collisions, see Fig.~\ref{fig:jpsi_sps_model} middle. Hadronic transport models like ultra-relativistic quantum molecular dynamics (UrQMD)~\cite{Spieles:1999kp} or hadron string dynamics (HSD)~\cite{Bratkovskaya:2003ux} also carried out a more rigorous description of the fireball evolution along with a dynamical treatment of the inelastic interactions between comoving hadrons and charmonia, by tracing the trajectory of each $J/\psi$ meson throughout the medium. However charmonium hadron cross section was still kept as an adjustable parameter in the model calculations. Efforts were also made to compute the charmonium disintegration cross section with light mesons either in quark or in hadronic models. 
$J/\psi$ suppression in Pb+Pb collisions at SPS was also found to be reasonably described by different variants of the QVZ approach~\cite{Bhaduri:2014daa,ch02,ch02b,Qiu:1998rz,Bhaduri:2012ta}, which treats the conventional nuclear matter absorption in an unconventional manner. Within this model scenario, heavy quark rescatterings in the cold nuclear medium increase the invariant mass of the evolving $c\bar{c}$ 
pairs such that some of the pairs originally produced within the mass interval between $2m_{C}$ to $2m_{D}$ can gain energy to cross the open charm threshold and transmute to $D\bar{D}$ pair causing  reduced $J/\psi$ production. Without leaving any room for additional suppression due to a hot and dense hadronic/partonic
medium created in the later stages of the collision, the model can reproduce the $J/\psi$ suppression pattern observed in central Pb+Pb collisions at SPS, as shown in Fig.~\ref{fig:jpsi_sps_model} right. 
At a fixed impact parameter, fluctuations of $E_{T}$ is found to play an important role in the interpretation of the data, particularly for central collisions. The solid line represents the ratio obtained without $E_{T}$ fluctuations and can reproduce the data upto the knee of the $E_{T}$ distribution. Model prediction gets saturated beyond the knee of the $E_{T}$ distribution whereas the data showed a rapid fall. The dashed line represents the model results incorporating the $E_{T}$ fluctuations, estimated following a Gaussian distribution of $E_{T}$~\cite{bl00} at a fixed impact parameter. \\

To obtain a coherent picture of charmonium production and suppression in heavy-ion collisions, NA50 complemented the $J/\psi$ studies by measurements of $\psi(2S)$ production using an identical analysis procedure as of $J/\psi$. Being a more loosely bound state than $J/\psi$, $\psi(2S)$ is likely to be melted at much lower temperature in a de-confined medium. In general, the measurement of $\psi(2S)$ is more challenging than $J/\psi$. Being more massive, $\psi(2S)$ has {\bf a} smaller production cross section than $J/\psi$. In addition, due to the smaller branching ratio in the dimuon decay channel and the stronger suppression in heavy-ion collisions, the extraction of the $\psi(2S)$ yield and the estimation of the associated systematic uncertainties are more challenging compared to the $J/\psi$ case. 
The study of $\psi(2S)$ production as a function of centrality in Pb+Pb collisions is shown in Fig.~\ref{fig:psi_prime_na50_PbPb}, by combining the data collected in the years 1998 and 2000 and using $E_{T}$ as the estimator of centrality. Results show a monotonic fall from peripheral to central collisions leading to a $\psi(2S)$ suppression by a factor of 6 relative to DY production and of 2.5 relative to $J/\psi$. Comparison with lighter collision systems indicates a stronger suppression of $\psi(2S)$ yield in Pb+Pb collisions than in p+A reactions~\cite{NA50:2006yzz}. Explicit comparison between the centrality dependent suppression patterns of $J/\psi$ and $\psi(2S)$ also shows that the anomalous suppression for $\psi(2S)$ sets in for smaller values of collision centrality, in line with the idea of sequential charmonium melting in a QGP medium. \\


\begin{figure}
  \includegraphics[width=1.0\linewidth]{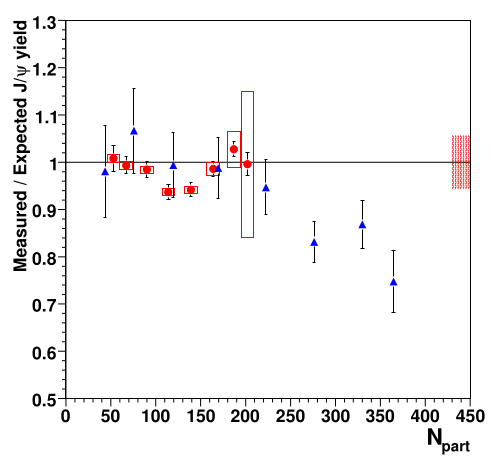}
\caption{ Ratio between measured and expected $J/\psi$ yields in 158 A GeV In+In (circles) and Pb+Pb collisions (triangles), as a function of $N_{part}$, as respectively measured by the NA60 and NA50 experiments at the SPS facility. The figure is adopted from~\cite{Arnaldi:2009ph}. Correlated systematic uncertainties are represented by boxes around the In+In data points. The filled box corresponds to the uncertainty on the absolute $J/\psi$ normalization of the In+In data points. The expected $J/\psi$ yield (reference for normal nuclear absorption) is estimated using Glauber model calculations with $\sigma_{abs}^{J/\psi} \sim 7.6$ mb. Uncertainty in $\sigma_{abs}^{J/\psi}$  leads to additional $12 \%$ global systematic error not shown here.}
\label{fig:na60_new}       
\end{figure}

The presence of anomalous $J/\psi$ suppression in Pb+Pb collisions and its absence in S+U collisions naturally motivated the investigation of an intermediate collision system to obtain further insights about this phenomena. Hence the NA60 experiment performed the measurement of $J/\psi$ suppression in In+In collisions at the same energy and kinematic domain of the NA50 heavy-ion collisions~\cite{NA60:2006ncq,Scomparin:2007rt,Scomparin:2010zz,Arnaldi:2009ph,NA60:2007odv,Arnaldi:2009xit}. Two different and complementary approaches have been adopted by the NA60 experiment to study $J/\psi$ suppression in nuclear collisions.  In the first approach, analogous to the method employed by the NA38 and NA50 experiments, the measured $J/\psi$ cross section was normalized to the DY yield in the di-muon mass window $2.9 < M_{\mu\mu} < 4.5$ GeV/$c^2$. Though the ratio has the advantage of being free from systematic uncertainties connected with beam luminosity and efficiency calculations, the small size of the sample of DY muon pairs collected by the NA60 experiment allowed to estimate the $\sigma_{J/\psi} \over \sigma_{DY}$ ratio only in three centrality bins. 
The corresponding results were seen to suffer from large uncertainties giving a clear indication that the use of DY data for normalization had to be avoided, for improved statistical significance. Hence an alternative approach was introduced by bypassing the DY process, where the measured $dN_{J/\psi}/dE_{ZDC}$ was directly compared to a reference spectrum estimated within a pure nuclear dissociation scenario. The statistical uncertainties associated with this second approach were small, around $2 \%$, whereas the systematic uncertainty was found out to be around $10 \%$ independent of centrality and essentially originating from the uncertainties in the Glauber model analysis of the nuclear absorption scenario. 
With the corresponding absorption cross section of $\sigma_{abs}^{G} \simeq 7.6$ mb, directly obtained from the $J/\psi$ production cross sections in 158 GeV p+A collisions, a reduction in the anomalous $J/\psi$ suppression in heavy-ion collisions is observed with respect to the previous estimate. Moreover while building the reference curve, the modification of nuclear parton densities was not explicitly taken into consideration. While in p+A collisions only parton distributions inside the target are affected due to shadowing, in A+A collisions parton distributions inside the projectile also get modified. This additional contribution needs to be accounted in the extrapolation of the CNM effects from p+A to A+A collision. Neglecting the shadowing effects in the extrapolation was found to introduce a small bias, generating a $\sim 5 \%$ artificial contribution to the suppression of the $J/\psi$ yield (for the EKS98 nPDF scheme). An additional reduction of the $J/\psi$ suppression was observed if shadowing corrections were suitably considered in the extrapolation from p+A to A+A collision data. The final ratio of the measured and expected $J/\psi$ yield is presented in Fig.\ref{fig:na60_new}, as a function of $N_{part}$ for both In+In and Pb+Pb collisions. The relative $J/\psi$ yield in In+In collisions shows small or no anomalous suppression and is compatible within  uncertainties with the CNM suppression. An additional $\sim 25 - 30 \%$ suppression still remained visible in very central ($N_{part} > 200$) Pb+Pb collisions, beyond the CNM effects. However the origin of the dip or "wiggle" at intermediate centrality for In+In collisions is still unknown. The quantum mechanical evolution of the quarkonium states under mutual interplay of absorption, screening and regeneration effects have been studied in~\cite{Pena:2013hd} within a time dependent harmonic oscillator model with complex oscillator strength. The wiggle has been predicted as an indication of the true threshold for anomalous suppression due to the coupling of $J/\psi$ to the $\bar{D}^{*}D^{0}$ channel with the $X(3872)$ state in the charmonium spectrum. \\

On passing we may briefly note that additional to $J/\psi$, NA60 also attempted to measure $\psi(2S)$ production in In+In collisions~\cite{Scomparin:2007rt}. Contrary to the $J/\psi$ detection, the significance of this measurement was limited by the small size of the data sample, consisting of about 300 $\psi(2S)$ mesons. Preliminary analysis of the ratio $\sigma_{\psi(2S)} \over \sigma_{DY}$ could be performed only in three centrality bins for In+In collisions and indicated that an anomalous suppression of $\psi(2S)$ sets in for more peripheral collisions than for the $J/\psi$ plausibly due its larger physical size and weaker binding, though a robust conclusion could not be reached due to its large statistical uncertainties. Apart from SPS, no other fixed target facility has so far measured charmonium production, in heavy-ion collisions. \\


\begin{figure*}
  \includegraphics[width=0.35 \linewidth]{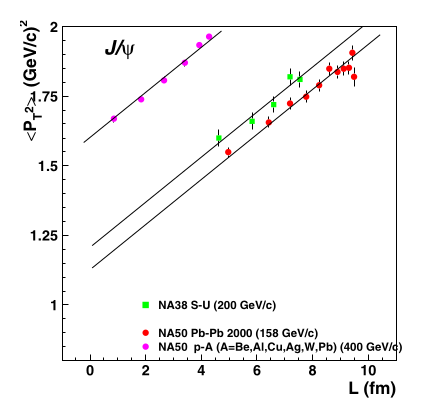}
  \includegraphics[width=0.35 \linewidth]{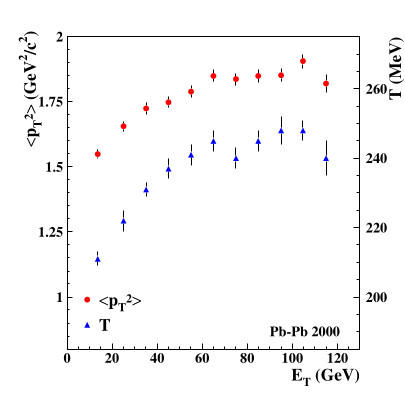}
  \includegraphics[width=0.35 \linewidth]{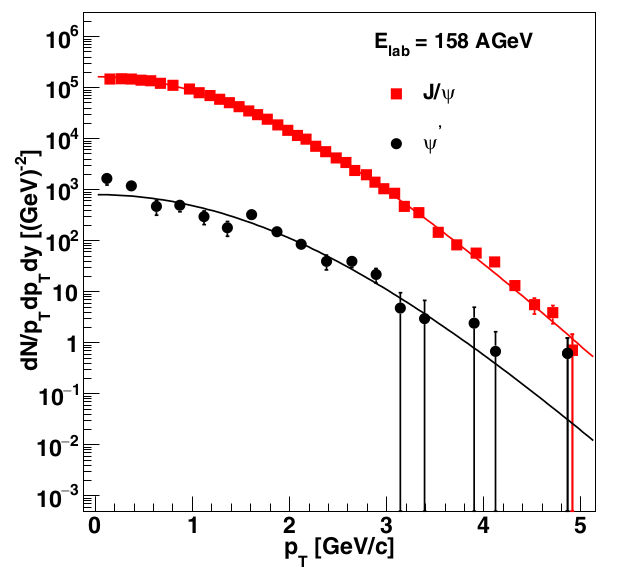}
\caption{(left) Variation of the average $p_{T}^{2}$ of $J/\psi$ mesons as a function of $L$ in p+A, S+U and Pb+Pb collisions measured by {\bf the} NA50 experiment at CERN-SPS facility. (middle) Mean $p_{T}^{2}$ and inverse slope parameter temperature ($T$) vs. L of $J/\psi$ mesons in 158 A GeV Pb+Pb collisions, as reported in~\cite{Ramello:2006db}. (right) Simultaneous fit to the $p_T$ spectra of $J/\psi$ and $\psi(2S)$ mesons following a non boost-invariant blast wave model~\cite{Rode:2020vhu} in 158 A GeV central Pb+Pb collisions.}
\label{fig:jpsi_na50_pT_T}       
\end{figure*}

Different important aspects of charmonium production dynamics in nuclear collisions can also be addressed through the kinematical distribution of the produced $J/\psi$ mesons. While multiple scattering of gluons in the initial state is expected to influence the $p_{T}$ distribution~\cite{Blaizot:1988hh}, the rapidity distribution may be sensitive to the underlying mechanism of $c\bar{c}$ hadronization~\cite{Vogt:1999dw}.
The $p_{T}$ distribution and the $p_{T}$ dependent suppression pattern of the produced charmonium states in 158 A GeV Pb+Pb collisions~\cite{NA50:2000mfb,Ramello:2006db} were measured by the NA50 Collaboration. {\bf $J/\psi$ ($\psi(2S)$)} $p_{T}$ distributions were studied in 11(5) centrality bins estimated in terms of $E_{T}$. For the $J/\psi$, the values of $<p_{T}>$ and $<p_{T}^{2}>$ were found to first increase from peripheral to central collisions and eventually  tend to flatten for more central collisions. The $<p_{T}^{2}>$ values were seen to be larger for $\psi(2S)$ than those of $J/\psi$. When plotted as a function of $L$ for p+A, S+U and Pb+Pb systems at different energies, the $<p_{T}^{2}>$ of $J/\psi$ mesons was seen to linearly rise with $L$, as illustrated in Fig.~\ref{fig:jpsi_na50_pT_T} left. This feature can be explained with initial state parton multiple scattering. For all the investigated systems the observed linear increment was found to be described using $<p_{T}^{2}>(L) = <p_{T}^{2}>_{pp} + a_{gN}L$, with an energy-dependent proton-proton coefficient $<p_{T}^{2}>_{pp}$ and a common slope parameter: $a_{gN} = 0.081 \pm 0.002$ GeV$^2$/c$^2$/fm. 
The transverse mass ($m_{T}$) distributions of the produced muon pairs were also studied in order to allow a comparison with thermal models. The extracted effective temperatures ($T_{eff}$) were obtained from a fit to the $J/\psi$ {\bf $m_{T}$} distributions in different centrality bins using
the analytical 
form $m_{T}^{2}K_{1}(m_{T}/T_{eff})$, where $K_{1}$ denotes the modified Bessel function of the second kind. The values of $T_{eff}$ were seen to first rise with $E_{T}$ and then eventually saturate similar to $<p_{T}^{2}>$ as it can be seen in the middle panel of Fig.~\ref{fig:jpsi_na50_pT_T}. A more rigorous analysis of $J/\psi$ and $\psi(2S)$ $p_{T}$ distributions in the blast wave model framework indicates that at SPS energies, these heavy mesons leave the medium immediately after the hadronization of the fireball~\cite{Gorenstein:2001ti,Rode:2020vhu}. The right panel of Fig.~\ref{fig:jpsi_na50_pT_T} shows a simultaneous fit to the $p_{T}$ spectra of $J/\psi$ and $\psi(2S)$ mesons within a non boost-invariant blast wave model framework. Due to their heavy mass, they have a small re-scattering cross section in the hadronic phase and thus exhibit a kinetic freeze-out temperature $T_{kin} \simeq 170$ MeV and an average transverse velocity $<\beta_{T}> \simeq 0.2$ and leave the expanding fireball much earlier than the bulk hadrons, which are known have much smaller $T_{kin}$ and larger radial flow~\cite{Rode:2018hlj}. 
To further investigate the $p_{T}$ dependence of the suppression pattern, the ratio, $F(E_{T},p_{Ti})$ defined as \\
$F(E_{T},p_{Ti}) = N_{\psi}(E_{T}, p_{Ti})/N_{DY}(E_{T})$ was calculated for 11 $J/\psi$ $p_{T}$ bins and 5 centrality bins in $E_{T}$. The denominator in the ratio, $N_{DY}(E_{T})$ denoted the $p_{T}$ integrated DY yield, proportional to the number of binary nucleon-nucleon collisions estimated for each $E_T$ bin. 
The results exhibited stronger suppression in the low $p_{T}$ bins with maximum suppression in the lowest bin. High $p_T$ bins have relatively flatter centrality dependence, indicating that anomalous suppression is more effective in the low  $p_{T}$ regime. For a finite-size fireball, such as the one produced in Pb+Pb collisions, low $p_{T}$ $J/\psi$ mesons have lower velocities and therefore spend more time inside the medium than the high $p_{T}$ $J/\psi$'s. Hence, one expects low $p_{T}$ $J/\psi$ mesons to be more affected by the medium effects and consequently more suppressed. Explicit evaluation of central-to-peripheral ratio, $R^{i}_{cp}(p_{T})$ in different centrality classes also indicated that anomalous $J/\psi$ suppression is prevalent at low $p_T$. For $p_{T} > 3.5$ GeV/c a very weak centrality dependence (if any) was observed. \\

\begin{figure*}
  \includegraphics[width=1.0 \linewidth]{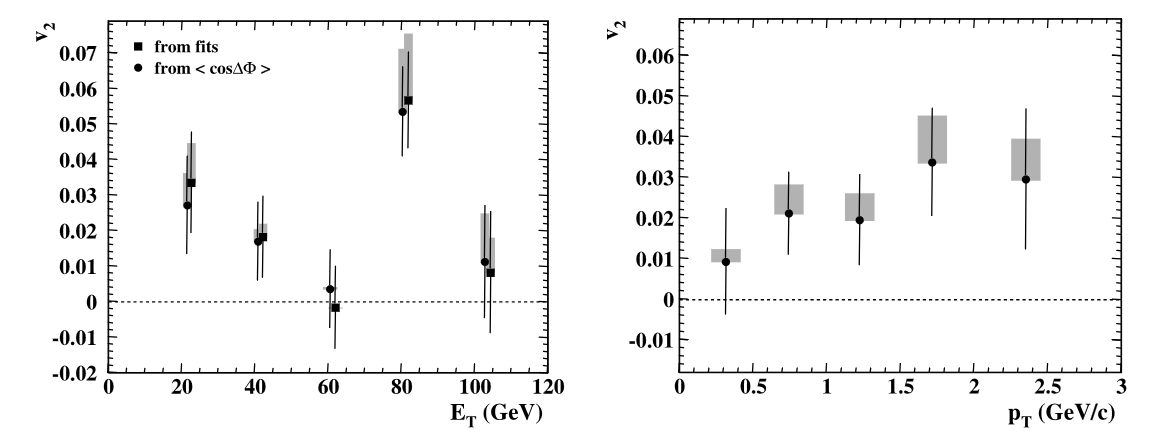}
\caption{Second Fourier coefficient $v_{2}$ of $J/\psi$ as a function of collision centrality, measured in terms of $E_{T}$ (left) and transverse momentum, $p_{T}$ (right) in 158 A GeV Pb+Pb collisions~\cite{NA50:2008lqf}. Statistical uncertainties are denoted by error bars and the systematic {\bf uncertainties} originating from the determination of event plane resolution are represented by the gray band.}
\label{fig:jpsi_na50_v2}       
\end{figure*}

The NA50 Collaboration also analyzed the $J/\psi$ azimuthal distribution relative to the reaction plane in Pb+Pb collisions using the data sample collected in year 2000~\cite{Prino:2007zz,Prino:2007ip,NA50:2008lqf}. For non-central heavy-ion collisions, the initial
geometrical anisotropy of the overlap zone between the projectile and target nuclei leads to an observable anisotropy in particle distribution in the momentum space, if the produced system interacts strongly enough to achieve thermalization at an early phase and build up collective flow. 
Hence, transverse anisotropic flow is expected to be found for $J/\psi$ mesons formed via $c-\bar{c}$ recombination if the re-scatterings are strong enough to lead to the (partial) thermalization of the charm quarks in the QGP\footnote{The charm flow is a "late" signal that builds up from a Brownian like motion of heavy quarks in the QGP, governed by the relaxation time of the heavy quarks.}. At SPS energies, where $c\bar{c}$ recombination is negligible and thermalization of primordial charm quarks is unlikely, charmonia are not expected to flow. However other sources related to azimuthally dependent $c\bar{c}$ dissociation processes either of partonic~\cite{Wang:2002ck,Zhu:2004nw} or of hadronic~\cite{Heiselberg:1999mf} origin are predicted to introduce an anisotropic azimuthal distribution of the observed $J/\psi$ mesons. Thus the measurement of the elliptic anisotropy, quantified by the second Fourier coefficient $v_2$ of the azimuthal distribution of $J/\psi$ relative to the reaction plane, serves as a useful probe for constraining the phenomenological models aimed at describing the observed anomalous $J/\psi$ suppression in Pb+Pb collisions. With this aim NA50 measured the normalized difference between the $J/\psi$ yields emitted in plane and out-of-plane as well as the second Fourier harmonic $v_{2}$. These quantities were measured  as a function of collision centrality (defined by $E_T$) and as a function of di-muon $p_{T}$. The orientation of the reaction plane was determined from the azimuthal anisotropy of $E_{T}$. Results for centrality dependence of $v_{2}$ are shown in the left panel of Fig.~\ref{fig:jpsi_na50_v2}. 
For non-central collisions, the extracted $v_{2}$ values were found to be significantly larger than zero indicating more $J/\psi$ emitted in-plane than out-of-plane, with maximum $v_{2}$ in the mid central bin, $70 < E_T < 90$ GeV, corresponding to an average impact parameter of 4.8 fm. 
Statistical uncertainties on the measurement of the $J/\psi$ anisotropy are denoted by
error bars. Systematic uncertainties are represented by the gray bands originating from the uncertainty on the determination of the event plane resolution. The $p_{T}$ dependence of the centrality integrated $J/\psi$ $v_{2}$ is illustrated in the right panel of Fig.~\ref{fig:jpsi_na50_v2}. \\



\begin{figure*}
  \includegraphics[width=0.33 \linewidth]{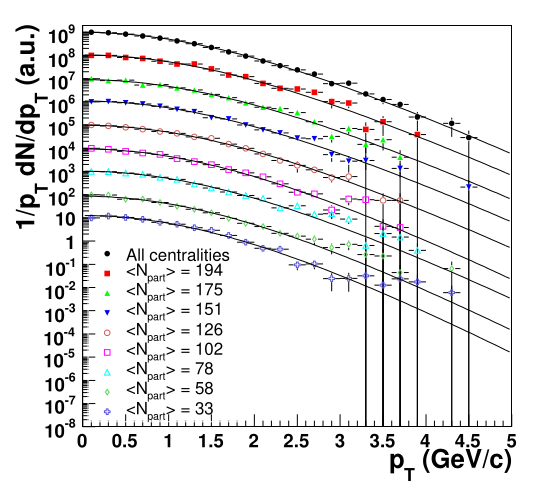}
  \includegraphics[width=0.33 \linewidth]{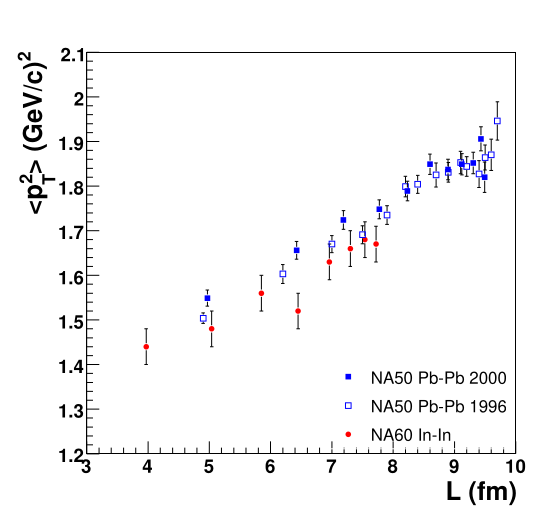}
  \includegraphics[width=0.33 \linewidth]{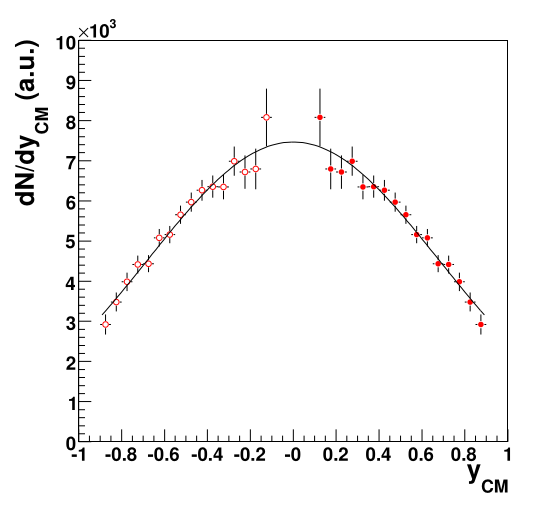}
\caption{(left) The acceptance corrected $p_{T}$ distribution of $J/\psi$ mesons produced in 158 A GeV In+In collisions. The lines correspond to the exponential fits ($\propto e^{-m_{T}/T}$) to the spectra in various centrality bins as well as centrality integrated spectra. (middle) The $L$-dependence of the $<p_{T}^{2}>_{J/\psi}$ in 158 A GeV In+In and Pb+Pb (two data sets corresponding to two different data taking periods) collisions. (right) Centrality integrated rapidity distribution of the $J/\psi$ mesons produced in 158 A GeV In+In collisions. Measured data points are represented by the closed symbols, which are reflected around $y_{cms}=0$ (open symbols) assuming a symmetric rapidity distribution. The figures are adopted from~\cite{NA60:2007odv}.}
\label{fig:jpsi_na60_kine}       
\end{figure*}

Alike NA50 experiment, NA60 also reported the kinematical distributions of the produced $J/\psi$ mesons in In+In collisions. The acceptance corrected $p_T$ spectra {\bf were} measured in the kinematic region corresponding to $0.1 < y_{cms} < 0.9$ and $-0.4 < cos(\theta_{H}) < 0.4$, integrated over centrality and for various centrality bins, as shown in Fig.~\ref{fig:jpsi_na60_kine} left\footnote{The angle in the Helicity frame, $\theta_{H}$, denotes the decay angle of the $\mu^{+}$, measured in the charmonium rest frame, with respect to the $J/\psi$ direction in the CMS frame.}. The measured ${1 \over p_{T}} dN/dp_{T}$ spectra were fitted with an exponential function of the form $e^{-m_{T}/T_{eff}}$, where $m_{T} = \sqrt{m^{2} + p_{T}^{2}}$, allowing an estimate of the effective temperature $T_{eff}$. The extracted temperature 
values were found to increase from 204 MeV to 234 MeV with increasing collision centrality. The centrality integrated value is $T_{eff} = 231 \pm 2$ MeV.
The $<p_{T}^{2}>_{J/\psi}$ was also found to exhibit a linear rise with the traversed length of nuclear matter, $L$, consistent with Pb+Pb collisions (see Fig.~\ref{fig:jpsi_na60_kine} middle), indicating  a centrality dependent $p_{T}$ broadening, a feature consistent with the occurrence of initial-state multiple scattering of the gluons as discussed above. The centrality integrated rapidity distribution obtained within the phase space domain $p_T <5$ GeV/$c$ and $|cos\theta_{H}| < 0.4$ was fitted by a symmetric Gaussian function, with $\sigma_{y} = 0.68 \pm 0.02$, as shown in Fig.~\ref{fig:jpsi_na60_kine} right. No significant dependence of $\sigma_{y}$ on collision centrality was observed.

\begin{figure*}
  \includegraphics[width=1.0 \linewidth]{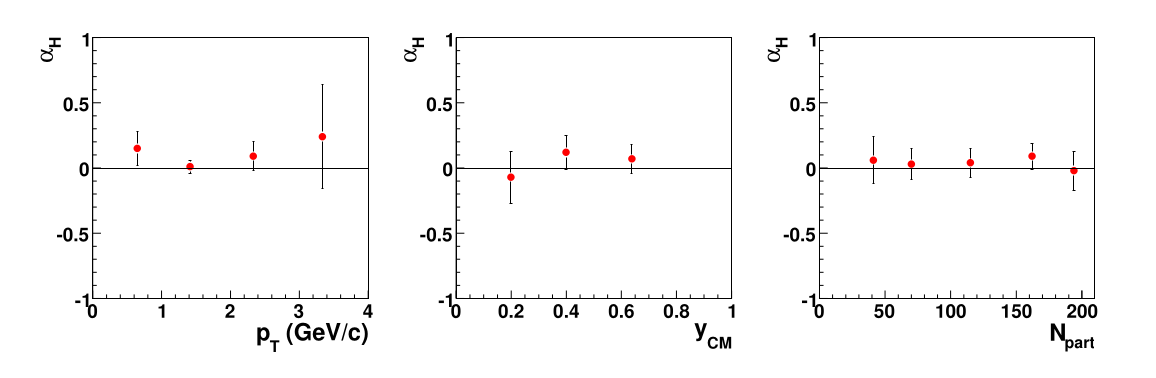}
\caption{The polarization of $J/\psi$ mesons produced in 158 A GeV In+In collisions, adopted from~\cite{NA60:2007odv}, as a function of $p_{T}$ (left), $y_{cms}$ (middle) and centrality expressed in terms of $N_{part}$ (right), measured by the NA60 Collaboration.}
\label{fig:jpsi_na60_pol}       
\end{figure*}


 $J/\psi$ polarization can be studied through the angular distribution of the decay muons. Different production models predict different polarization of the $J/\psi$ mesons, which may also be affected by the deconfined medium~\cite{Ioffe:2003rd}. The NA60 experiment also reported first measurements of $J/\psi$ polarization in heavy-ion collisions.  To estimate the polarization parameter, the angular distributions of the decay muons were fitted
using the functional form  ${d\sigma \over d cos\theta_{H} } = 1 + \alpha_{H} cos^{2} \theta_{H}$ , where $\theta_{H}$ is the polar decay angle of the $\mu^{+}$ in the Helicity reference frame. A transverse polarization of the produced $J/\psi$ mesons would result in $\alpha_{H} > 0$, whereas $\alpha_{H} < 0$ indicates longitudinal polarization. As shown in Fig.~\ref{fig:jpsi_na60_pol} the $J/\psi$ mesons produced in In+In collisions do not show any significant polarization as a function of centrality, $y$ and $p_T$ corresponding to $\alpha_{H}$ compatible with zero within uncertainties.\\ 
\begin{figure*}
  \includegraphics[width=1.0 \linewidth]{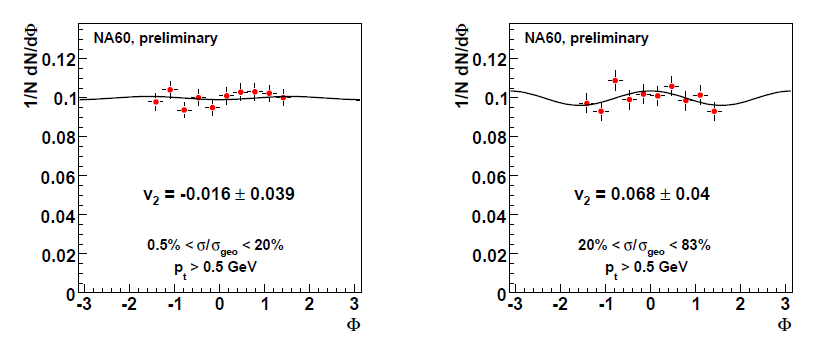}
\caption{Azimuthal distribution of $J/\psi$ produced in 158 A GeV In+In collisions for (left) central and (right) peripheral events. The corresponding values of elliptic anisotropy, $v_{2}$ estimated via the event plane method are also shown, after correcting for the event plane resolution, as reported in~\cite{NA60:2007odv}.}
\label{fig:jpsi_na60_v2}       
\end{figure*}
The azimuthal distribution of $J/\psi$ mesons produced in In+In collisions were also reported by the NA60 experiment. The distributions were measured with respect to the orientation of the reaction plane determined via the event plane method, from the emission angles of charged particles registered in the vertex tracker. Fig.~\ref{fig:jpsi_na60_v2} displays the azimuthal angle distribution of $J/\psi$ for (left) central and (right) peripheral collisions along with the corresponding $v_2$ estimated after correcting for the event plane resolution. Indication of non-isotropic emission pattern was seen for more peripheral events but was not conclusive due to the large statistical uncertainties. \\

\section{Prospects for measurement of charmonium production below top SPS energy}
\label{sec:3}

\begin{figure*}
  \includegraphics[width=1.0\linewidth]{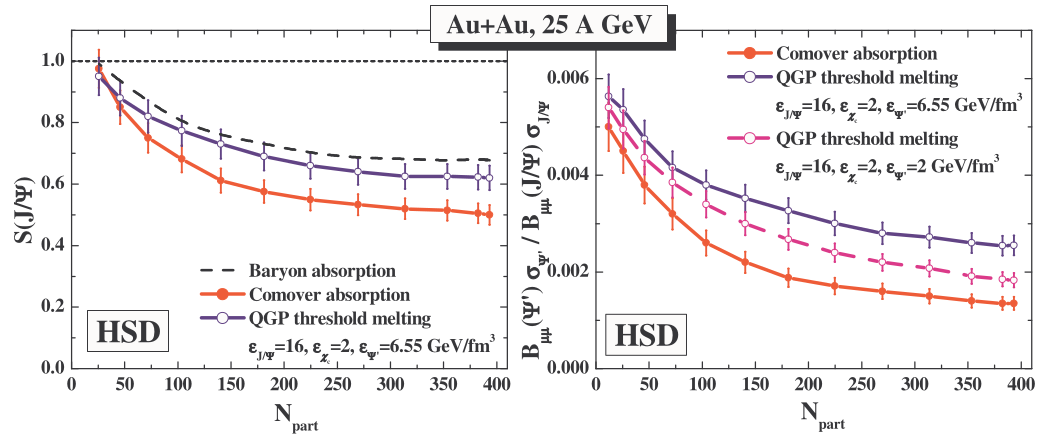}
\caption {$J/\psi$ survival probability, $S_{J/\psi}$ (left panel) and $\psi(2S)/ J/\psi$ ratio in the di-muon decay channel (right panel) as a function of $N_{part}$ in 25 A GeV Au+Au collisions, as calculated by the HSD transport model. The figure is taken from~\cite{Linnyk:2006ti}. The violet lines (with open dots) mimic the QGP threshold melting scenario for critical energy densities, $\epsilon_{\psi^{'}} = 6.55$ GeV/fm$^3$, $\epsilon_{\chi_{c}} = 2$ GeV/fm$^3$ and $\epsilon_{J/\psi} = 16$ GeV/fm$^3$. The magenta line (the lower line with open dots on the right panel) stands for the threshold melting scenario with $\epsilon_{J/\psi} = 16$ GeV/fm$^3$, $\epsilon_{\chi_{c}} = \epsilon_{\psi^{'}} = 2$ GeV/fm$^3$. The solid red lines (full dots) denote the suppression due to hadronic comovers. The dashed line (left panel) represents the HSD calculations including only dissociation in cold nuclear medium.}
\label{fig:suppression_fair_hsd}       
\end{figure*}

 As discussed in the previous section comprehensive experimental measurements of charmonium production, particularly focusing on the $J/\psi$ meson, have been carried out in p+A and A+A collisions at the SPS. The most intriguing results obtained from SPS are possibly the discovery of anomalous $J/\psi$ suppression beyond the cold matter effects and the first hint of sequential charmonium suppression, both observed in central 158 A GeV Pb+Pb collisions. What additional light can such studies shed if measurements are performed at lower collision energies? 
  At lower $\sqrt{s_{NN}}$, lower temperatures are attained in the fireball, resulting in reduced dissociation/melting of the primordially produced charmonium states. In addition, the lower medium density and shorter QGP lifetime reduce the interactions of the charm quarks in the medium, making it more difficult for them to reach (partial) thermalization in the QGP. This may in turn affect the endogamous recombination probability (exogamous recombination is already negligible at top SPS energy, as discussed above). Hence, the effects of melting/dissociation and medium interactions are expected to depend on the collision energy. Note that these considerations are to some extent qualitative, and a more quantitative picture should be provided by phenomenological models including charm quark transport in a hydrodynamically expanding medium in low energy collisions. \\

 Similar to 158 A GeV beam energy, directly produced $J/\psi$ mesons, owing to their strong binding energy are not expected to undergo significant suppression by the deconfined medium, in low energy regime. However higher mass charmonium states ($\psi(2S)$, $\chi_{c}$) being weakly bound may still get dissociated leading to the suppression of their yields if they can be detected directly or may modify the measured $J/\psi$ yield due to the suppression of the feed-down from their decays. As the collision energy decreases, leading to progressively lower initial temperatures, dissociation effects on $\chi_{c}$ and $\psi(2S)$ are also expected to diminish. By performing a beam energy scan of $J/\psi$ suppression in heavy-ion collisions it might at least in principle be possible to identify the threshold beam energy at which dissociation of these states begins. Such a measurement of $J/\psi$ production as a function of collision energy will also be useful to complement the studies performed so far at top SPS energy as a function of collision centrality.

  \begin{figure}
  \includegraphics[width=1.0\linewidth]{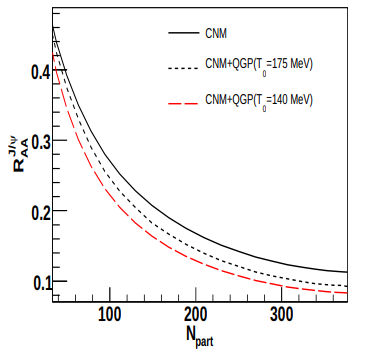}
\caption{Model prediction for the variation of $R_{AA}^{J/\psi}$ as a function of $N_{part}$ in 30 A GeV Au+Au collisions. Results are obtained from~\cite{Bhaduri:2013uoa}. $J/\psi$s are seen to be dominantly suppressed by the CNM effects. Formation of the deconfined matter enhances the dissociation, maximum by 15 to 25 $\%$ depending on the opted value of $T_{0}$, an input parameter of the underlying plasma EoS.}
\label{fig:suppression_fair_qgp_ppb_2013}       
\end{figure}

 Correlating the threshold with precise measurements of the fireball temperature from the invariant mass distribution of the intermediate mass ($1.5 < M_{\mu^{+}\mu^{-}} < 2.5 $ GeV/$c^2$) thermal dimuons~\cite{Usai:2024row} might help to estimate experimentally the critical temperature at which these charmonium states dissolve. This provides a crucial test for theoretical predictions of melting temperatures from Lattice QCD calculations. To date, the modeling of charmonium dynamical evolution in dense baryonic matter, such as the one produced in low energy nuclear collisions, has not been extensively investigated in literature. Recently there have been attempts to model the quarkonium suppression originated in the pre-equilibrium glasma phase, prior to the formation of QGP in heavy-ion collisions at the LHC energy domain~\cite{Pooja:2024rnn}. Impact of such pre-equilibrium dissociation is expected to be more severe at lower energies due to the large formation time of the plasma phase. The dissociation process may also get influenced in a high $\mu_{B}$ medium due to the presence of excess valence quarks. Moreover, comover interactions quantifying inelastic interactions within the hadronic phase are expected to be more relevant at lower collision energies, due to the shorter life time of the plasma as compared to the hadronic stage. One of the early predictions on the charmonium suppression at low bombarding energies is available from the Hadron String Dynamics (HSD) transport model approach~\cite{Linnyk:2006ti}. Two different dissociation scenarios namely the QGP motivated threshold melting scenario and the hadronic comover scenario are seen to result in distinguishably different suppression patterns, in 25 A GeV Au+Au collisions, as shown in Fig.~\ref{fig:suppression_fair_hsd}. The centrality dependence of the $J/\psi$  survival probability and the $\psi(2S) \over J/\psi$ ratio are significantly lower in the comover absorption model. Owing to smaller initial energy densities reached in the fireball produced in low energy nuclear collisions, very small additional suppression is induced due to QGP in this model, as compared to normal nuclear absorption.
 
\begin{figure}
  \includegraphics[width=1.0\linewidth]{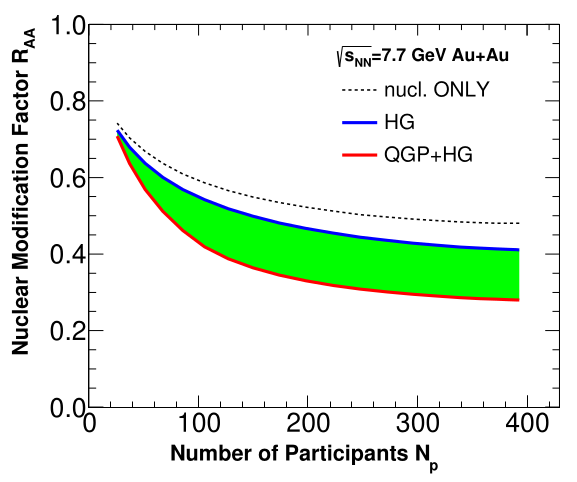}
\caption{Centrality dependence of the $J/\psi$ nuclear modification factor ($R_{AA}$) in $\sqrt{s_{NN}}= 7.7$ GeV Au+Au collisions, as calculated in~\cite{Chen:2015ona}. The black dotted line incorporates only normal nuclear absorption. Combined suppression  effects to due QGP and hadronic phases are shown by the red solid line.The blue solid line assumes no QGP formation and charmonium suppression induced only by norml nuclear absorption and interactions with comoving hadrons.} 
\label{fig:chen}       
\end{figure}


 \begin{figure*}
  \includegraphics[width=1.0\linewidth]{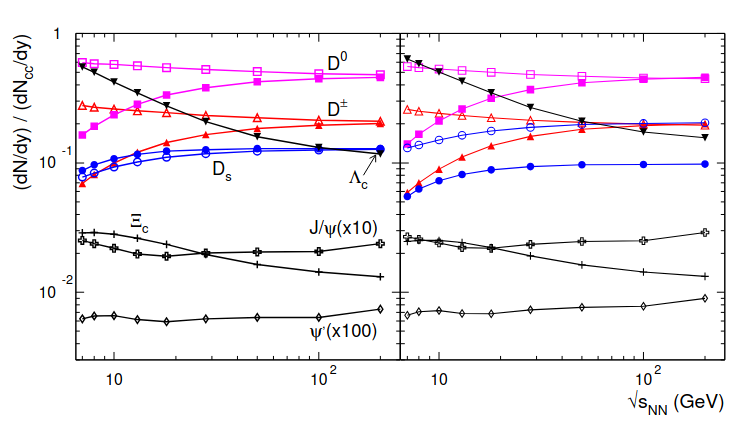}
\caption{Collision energy dependence of the charmed hadron yields relative to the charm quark pair yield for two theoretically predicted scenarios of the in-medium mass shift ($\Delta m$) of open charm hadrons made of light $u$ or $d$ quarks: (i) the left panel represents a common drop of $\Delta m = 50$ MeV for all charmed mesons and a drop of 100 MeV for $\Lambda_{c}$
and $\Sigma_{c}$ baryons (50 MeV decrease for $\Xi_{c}$); (ii) the right panel represents the second scenario where a decrease of $\Delta m=100$ MeV is assumed for all charmed mesons and a 50 MeV enhancement for their antiparticles. For charmed baryons, the same (scaled with the number of light quarks) scenario as
in (i) is assumed. For the $D$ mesons, the full and open symbols respectively represent particles and antiparticles. Vacuum masses are considered for $D_{s}$ mesons and charmonia in both scenarios. Yields of $J/\psi$ and $\psi(2S)$ are multiplied by factors of 10 and 100 respectively.} 
\label{fig:shm}       
\end{figure*}
 
  The survival probability of different charmonium states in an expanding baryon rich QGP, such as the one produced in central heavy-ion collisions at low bombarding energies, has been estimated in~\cite{Bhaduri:2013uoa}. The time evolution of the medium is incorporated following the UrQMD model which describes the full 3-D expansion of the fireball. Screening effects remain operative on a resonance state over the time it spends inside the plasma. A particular resonance state is assumed to undergo complete dissociation by color screening, if its spatial size becomes comparable to the in-medium Debye radius. The magnitude of suppression exhibits a strong dependence on the medium expansion dynamics. The presence of transverse expansion leads to faster cooling of the plasma leading to weaker suppression effects~\cite{Bhaduri:2015kxa}. Fig.~\ref{fig:suppression_fair_qgp_ppb_2013} shows the centrality dependence of $R_{AA}^{J/\psi}$ in 30 A GeV Au+Au collisions. The overall suppression pattern is overwhelmingly dominated by the CNM effects described with an adapted version of the QVZ model~\cite{Bhaduri:2012ta} that includes the nuclear modification of the parton densities implemented with \linebreak 
  EPS09~\cite{Eskola:2009uj} nPDF scheme and the multiple scattering of the initially produced $c\bar{c}$ pairs inside the nuclear medium. Debye screening may induce additional $15 - 20 \%$ suppression for the most central collisions, depending on the chosen value of $T_{0}$, an input parameter of the underlying plasma equation of state (EoS). As proposed by Kapusta~\cite{Kapusta:2010ke}, the EoS is constructed such that it reproduces the known properties of the ground state nuclear matter as well as the lattice QCD simulation results at zero net baryon density and $T_{0}$ can be identified with the pseudo-critical temperature ($T_{c}$) at $\mu_{B} \simeq 0$. In Ref.~\cite{Chen:2015ona} the authors have calculated the $J/\psi$ suppression in a hot and dense medium both within QGP and hadron gas scenarios. The medium evolution is modeled using (2+1) dimensional ideal hydrodynamics with the conservation equation of net baryon density. The time scale of the medium reaching local thermal equilibrium has been taken to be $\tau_{0} = 1$ fm/$c$ and a first order phase transition is assumed between QGP and hadronic phase. A Blotzmann-type transport equation is employed to model the phase space dynamics of the charmonium states.
 Suppression inside the plasma phase is modeled via inelastic collisions with hard gluons ($g + J/\psi \rightarrow c+\bar{c}$) whereas hadronic dissociation is mostly induced by $\pi$ and $\rho$ mesons. Charmonium regeneration due to recombination of $c\bar{c}$ or $D\bar{D}$ pairs is found to be negligible at low colliding energies. The centrality dependence of $R_{AA}^{J/\psi}$ evaluated under different dissociation scenarios in 25 A GeV ($\sqrt{s_{NN}} =7.7$ GeV) Au+Au collisions is shown in Fig.~\ref{fig:chen}. The black dashed line represents the suppression exclusively due to nuclear absorption suffered by charmonium when produced in the initial colliding times. The two solid lines represent two possible scenarios of anomalous suppression (in addition to normal nuclear absorption): one where the dissociation comes solely from the hot and dense hadronic gas phase assuming no formation of QGP (upper blue line), and the other where it originates from the QGP phase followed by a first order transition to the hadronic phase (lower red line). 
 The green band highlights the effect of the assumption regarding the QGP formation at such low collision energies.  
 These calculations have been further augmented in~\cite{Zhao:2022cvl}. 
Using MUSIC~\cite{Schenke:2010nt}, the (3+1)D viscous hydrodynamical model for the space time evolution of the bulk medium, the centrality dependence of $J/\psi$ $R_{AA}$ is estimated in the high baryon density medium produced in Au+Au collisions at energies of RHIC beam energy scan, namely, $\sqrt{s_{NN}}$ = 39, 14.5 and 7.7 GeV. The contribution due to finite $\mu_B$ is encoded in the heavy quark potential through the Debye screening mass and also in the EoS driving the expansion of the baryonic plasma. The effect of finite baryon density on the $J/\psi$ production is found to be negligible at $\sqrt{s_{NN}} = 39$ and 14.5 GeV. However a stronger suppression induced by finite $\mu_B$ is observed at $\sqrt{s_{NN}} = 7.7$ GeV, leading to a reduction in the $R_{AA}$ as a function of collision centrality. Experimental data in low energy nuclear collisions will thus be highly welcome to better constrain the charmonium evolution dynamics through a high $\mu_{B}$ medium. \\

Charmonium production in low energy nuclear collisions from near threshold ($\sqrt{s_{NN}} =6$ GeV) up to top RHIC energy has also been calculated within the framework of the SHMc model~\cite{Andronic:2007zu}. The yield of charmed baryons ($\Lambda_{c}$, $\Xi_{c}$) relative to the total $c\bar{c}$ yield is found to increase strongly with decreasing collision energy, because of the dominance of $\mu_{B}$ coupled with the condition of charm neutrality. Due to the large charm quark mass and a clear separation of timescales for charm quark and charmed hadron production, possible in-medium mass shifts of charmed hadrons in a hot and dense hadronic medium do not lead to measurable changes in the D-meson production cross sections, if charm conservation is explicitly taken into account. However, such medium effects would lead to a redistribution of charm quarks, which will influence the production rate of charmonia relative to $D$ mesons. For dropping in-medium masses, $D$-mesons will eat away the charm quarks of charmonia, which will be reflected in their excitation function, as shown in Fig.~\ref{fig:shm}, for two possible scenarios of in-medium mass shifts of open charm hadrons. These effects are visible at all beam energies and slightly rise close to threshold. \\

\begin{figure}
  \includegraphics[width=1.1\linewidth]{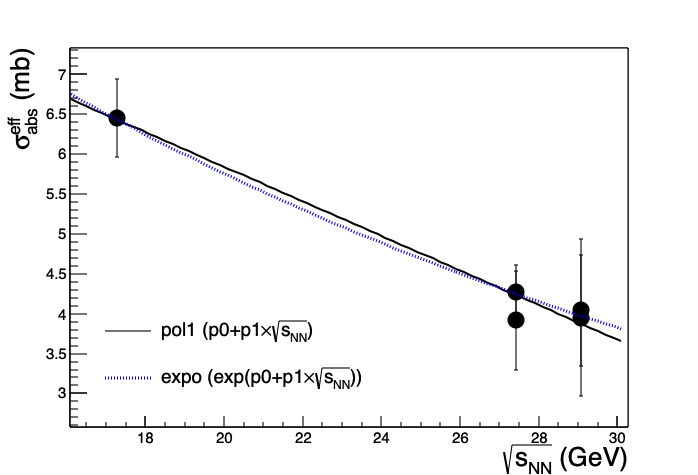}
\caption{Collision energy dependence of the effective nuclear absorption cross section ($\sigma_{abs}^{eff}$) in the SPS energy domain. Data are fitted using a first ($1^{st}$) order polynomial as well as an exponential function. The figure is adopted from~\cite{Chatterjee:2022ssu}.} 
\label{fig:sayak}       
\end{figure}

One fundamental aspect of studying charmonium production in nuclear collisions involves the understanding of the precise role of CNM effects. As previously discussed, at SPS there is already strong experimental evidence of enhanced CNM suppression in p+A collisions at lower incident proton beam energies. It is certainly important to extend these p+A collision measurements below the beam energy of 158 GeV. As seen in the model calculations, suppression effects due to a parton plasma are expected to diminish at lower collision energies due to reduced energy density and shorter life time of the produced medium. The magnitude of CNM effects is likely to get stronger with decreasing beam energy as the two colliding nuclei would take longer time to pass through each other. An accurate quantification of CNM effects thus becomes imperative in this regime.  As these effects are independent of QGP formation, they need to be properly accounted for to look for any hint of "anomalous" suppression in central heavy-ion collisions. In addition to constituting the suitable baseline for the corresponding heavy-ion measurements, the studies of charmonium production in cold QCD matter are interesting in their own right offering many physics insights hitherto unexplored at the existing facilities. Recording $J/\psi$ data in p+A collisions with varying proton beam energy is thus essential for low energy measurements. Due to the complex interplay among different  physics processes, a complete and unified interpretation of the existing p+A data is not yet possible. New measurements at lower collision energies, particularly in specific kinematic domains, can provide crucial constraints on the nature and magnitude of these effects, useful to disentangle their different contributions. A theory agnostic data driven approach has been adopted in~\cite{Chatterjee:2022ssu} to estimate the level of normal nuclear absorption in p+A collisions in the energy range between 30 to 158 GeV. The available p+A data on $J/\psi$ production from the NA50 and NA60 Collaborations are analyzed within a pure Glauber model ansatz based on the $\rho L$ parametrization for this purpose. The $\sqrt{s}$ dependence of the extracted $\sigma_{abs}^{eff}$ values are parametrized using a first order polynomial and also an exponential function as displayed in Fig.~\ref{fig:sayak}. Both energy dependent parameterizations are found to  have comparable fit qualities and when extrapolated to lower energies generate an absorption cross section in the range between 7 to 9 mb. An energy independent absorption scenario is found to be unlikely, as indicated by large $\chi^{2}$ and very small fit probability.\\
 


The foreseen measurement of $J/\psi$ production in p+A collisions below 158 GeV by the NA60+ Collaboration provides a unique possibility to detect the presence of an intrinsic charm component in the nucleon wave function~\cite{Vogt:2021vsc,Vogt:2022glr,Vogt:2023plx}. This non-perturbative charm production was first speculated in 1980s~\cite{Brodsky:1980pb,Brodsky:1981se,Brodsky:1989ex} and subsequently explored by several experiments~\cite{EuropeanMuon:1981obg} but its existence is yet to be satisfactorily confirmed. Presence of intrinsic charm is believed to result in an enhanced charm production at large $x_F$ because the heavy quarks carry a larger fraction of the projectile momentum than the light quarks in these states. 
Recent measurements by LHCb reported an excess of Z +c-jets over Z+jets alone at forward rapidity~\cite{LHCb:2021stx}, in agreement with a $1 \% $ intrinsic charm contribution in the proton. On the other hand, analysis with SMOG gas-jet target in the LHCb detector~\cite{LHCb:2018jry} claimed to see no evidence of intrinsic charm contribution in the $J/\psi$ or $D^0$ production. At collider energies the presence of intrinsic charm is expected to appear at extremely high values of forward rapidity and thus inaccessible by the experimental measurements. In fixed target experiments at lower energies, the predicted enhancement in charm production occurs closer to mid-rapidity. Model calculations~\cite{Vogt:2022glr} of the $J/\psi$ nuclear modification factor in p+A collisions at low SPS energies, assuming production from a combination of perturbative QCD and $|uudc\bar{c}>$ Fock state suggest that even a modest intrinsic charm component, at the level of $0.1 \%$, could lead to a dominant contribution to the observed production. \\

\begin{figure}
  \includegraphics[width=1.0\linewidth]{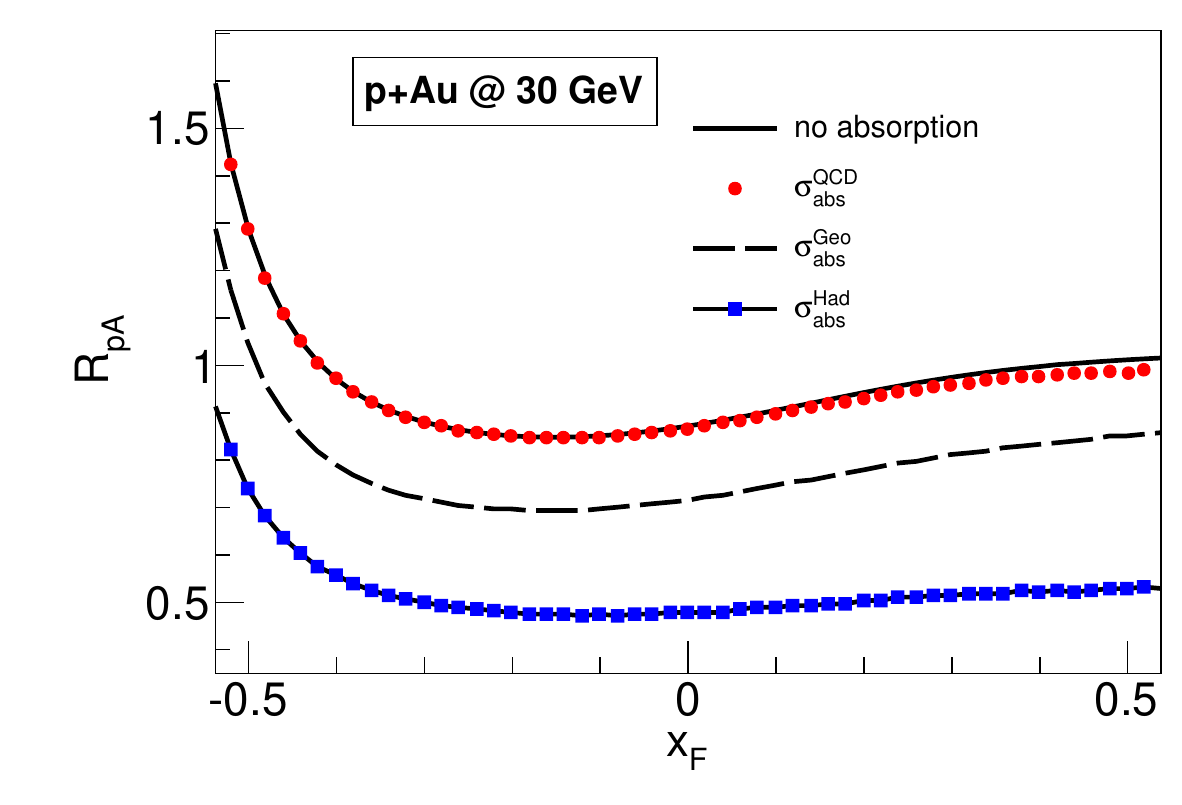}
\caption{Variation of $J/\psi$ $R_{pA}$ as a function of $x_F$ in 30 GeV p+Au collisions, adopted from~\cite{Bhaduri:2017ptw}. Different curves correspond to different models of resonance-nucleon inelastic collisions resulting in widely varying dissociation cross section in the FAIR energy domain. This ultimately leads to distinguishably different $J/\psi$ suppression patterns, as a function of $x_F$.}
\label{fig:RpA_jpsi_fair}       
\end{figure}

Measuring differential $J/\psi$ production in p+A collisions at FAIR SIS100 energies might be particularly interesting to probe resonance-nucleon interaction. The intrinsic formation time of $J/\psi$ mesons is around $\tau_{J/\psi} \simeq 0.35$ fm~\cite{Kharzeev:1995br}. In the laboratory frame, this formation time will be dilated and the corresponding Lorentz factor $\gamma$ will depend on the velocity of $J/\psi$ mesons in the rest frame of the target nucleus. Slow $J/\psi$ mesons will form early having typical formation length smaller than the size of the target nucleus whereas fast $J/\psi$ will form late outside the target. At FAIR, since $J/\psi$ production will occur near the kinematic threshold, the slowest possible $J/\psi$ mesons will be formed with an average laboratory formation length lower than 2 fm\footnote{See Fig.1 of Ref.~\cite{Bhaduri:2017ptw} for the $x_{F}$ dependence of the $J/\psi$ formation length in the nuclear rest frame, in 15 GeV and 30 GeV p+A collisions.}. For a heavy nuclear target, instead of a $c\bar{c}$ pair, a full grown resonance will thus propagate through the nuclear medium. Existing hadronic models predict widely varying magnitude of inelastic $J/\psi+N$ reaction cross section, none of which can either be accepted or rejected due to unavailability of measurements for suitable comparison. The various scenarios of $J/\psi$ dissociation, based on different theories of $J/\psi +N$ interaction in a cold medium in this energy regime, predict a very distinct survival probability leading to distinguishably different patterns of the so called transparency ratio, $R_{pA}$, as a function of $x_{F}$, as shown in Fig.~\ref{fig:RpA_jpsi_fair} for 30 GeV p+Au collisions~\cite{Bhaduri:2017ptw}. Measurement of $J/\psi$ production as a function of rapidity (or $x_{F}$) will thus be interesting to study the resonance propagation through a baryonic medium. The data recorded at higher energies, where the experimental phase space domain is predominantly populated by the propagation of the initially produced $c\bar{c}$ pairs through the nuclear matter is not very useful for this purpose~\cite{Bratkovskaya:2003ux}. Such an experimental discrimination of various absorption scenarios not only puts constraints on the corresponding theoretical models but also will be useful for the heavy-ion data collected in this regime.  Inclusive $J/\psi$ production in p+A collisions, in the near-threshold energy region has also been studied with a first collision model based on the nuclear spectral function~\cite{Kiselev:2014dza,Paryev:2017tmo}. Charmonia are assumed to be produced in incoherent primary p+N collisions and influenced by the nuclear $J/\psi$ mean-field potential. Under different scenarios of in-medium modification, the inclusive as well as differential $J/\psi$ production cross sections are studied for p+C, p+W and p+Au collisions at incident proton kinetic energies of 8, 11 and 23 GeV. Momentum distributions (absolute and relative) of the soft $J/\psi$ mesons produced at such low collision energies are found to exhibit distinct sensitivity to the $J/\psi$ in-medium modification, an effect that can be directly verified once the FAIR SIS100 accelerator becomes operational. \\

\begin{figure}
  \includegraphics[width=0.9\linewidth]{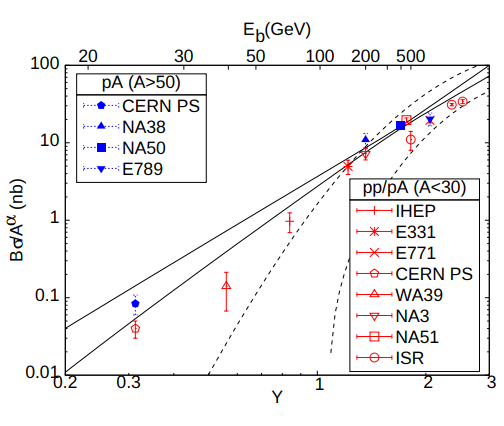}
\caption{Scaling plot for the available data on inclusive $J/\psi$ production cross section in p+A collisions. The band enclosed between the solid lines represents the $68 \%$ confidence limit of the two parameter fit following the scaling theory for $A > 50$. The band enclosed by the dashed lines is the 68 $\%$ confidence limit of a two parameter fit to the form $K{(1 - e^{-Y})}^{\nu}$. The figure is adopted from~\cite{Bhaduri:2013ica}}
\label{fig:jpsi_scaling}       
\end{figure}
A phenomenologically motivated scaling theory with two exponents $\beta$ and $\alpha$, has also been proposed in literature~\cite{Bhaduri:2013ica} to describe the near-threshold production of quarkonium resonances in cold nuclear matter, without assuming QCD factorization. 
The exponent $\beta$ accounts for the scaling of the ground-state quarkonium production cross section with collision energy (expressed in terms of $Y$\footnote{$Y$ is a dimensionless measure of the collision energy in the centre-of-mass (cms) frame and defined as $Y=ln\sqrt{s/s_{0}}$, where $\sqrt{s_{0}}$ is the cms threshold energy for $J/\psi$ production in p+p collisions.}) in p+p collisions. The exponent $\alpha$ characterizes the scaling with the nuclear mass number A in p+A collisions, and is introduced to model cold nuclear matter effects. In this model, $\alpha$ is assumed to depend linearly on $Y$.
The values of $\beta$ and $\alpha$ are fixed by performing a scaling analysis of near threshold $J/\psi$ production in p+p and p+A collisions. The multi-fractal exponent $\alpha$ is found to be extendable to describe the production of other mesons in p+A collisions. The limited data corpus currently available on $J/\psi$ production in p+A collisions has been found insufficient to provide a very stringent test of scaling in near-threshold production cross sections, as illustrated in Fig.~\ref{fig:jpsi_scaling}. The FAIR SIS100 facility presents an opportunity to probe the region $Y \le 0.4$ very thoroughly with modern statistics and thus  to test and refine such novel approaches of charmonium production.\\

\begin{figure}
\includegraphics[width=1.0\linewidth] {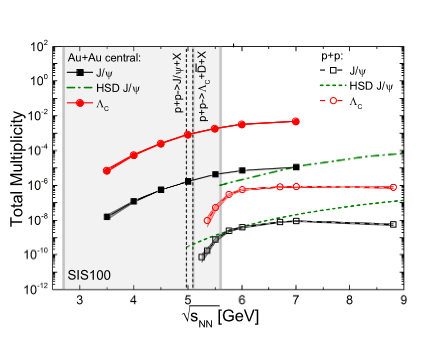}
\caption{Multiplicity of $J /\psi$ meson and $\Lambda_{c}^{+}$ baryon in p + p and central ($0 - 10 \%$) Au+Au collisions as a function of the collision energy, as obtained by the UrQMD model incorporating mechanisms for sub-threshold charm production. Results are compared with previous HSD model predictions. Vertical lines indicate the kinematic threshold energies of the corresponding channels in p + p reactions The vertical grey band represents the foreseen beam energy range for heavy-ion collisions at FAIR SIS100 accelerator. The figure is adopted from~\cite{Steinheimer:2016jjk}.}
\label{fig:subthreshold_jpsi}       
\end{figure}

\begin{figure}
\includegraphics[width=1.0\linewidth]{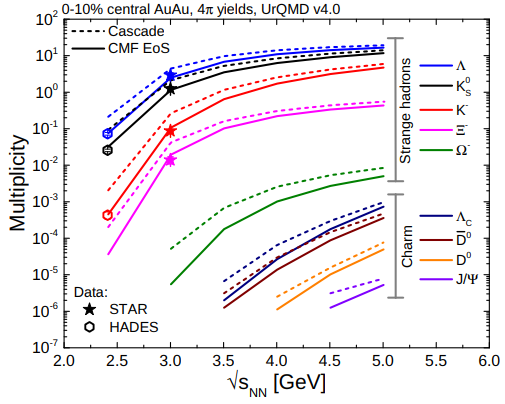}
\caption{Multiplicity of strange and charmed hadrons as a function of collision energy in central ($0 - 10 \%$) collisions. UrQMD model calculations (lines) are compared to the available data (symbols) from STAR BES and HADES experiments. Simulations with CMF EoS are denoted by solid lines, whereas dashed lines indicate the cascade model calculations. The figure is taken from~\cite{Steinheimer:2025trr}.}
\label{fig:subthreshold_jpsi_new}       
\end{figure}

The maximum beam energy that will be available from FAIR SIS100 accelerator is 11 A GeV, which is below the threshold for $J/\psi$ production in elementary collisions ($E_{b}^{th} = 12.5$ A GeV or equivalently $\sqrt{s_{NN}} = 4.97$ GeV) for the reaction $p + N \rightarrow J/\psi + p +N$). Thus charmonium production should be energetically forbidden in heavy-ion collisions at FAIR SIS100. However theoretical calculations based on the UrQMD model predict a novel mechanism of sub-threshold charm production in nuclear collisions at SIS100 energy domain~\cite{Steinheimer:2016jjk}. The production mechanism is driven by multi-step scatterings of nucleons and their resonance states, accumulating sufficient energy for the production of $J/\psi$ and other charmed hadrons in low energy collisions that will be available at FAIR. At higher collision energies charm production usually follows the so called binary scaling where all charm is essentially produced in the first binary collisions between nucleons of projectile and target nuclei. With decreasing the energy of the collision, the strict binary scaling is expected to undergo breakdown as the energy available per individual p + p collision is not kinematically sufficient for producing a $c\bar{c}$ pair. For particle production to occur at and below the elementary production threshold, secondary baryon interactions are thus essential. In the FAIR SIS100 energy domain, the produced system is likely to be close to the transition between the hadronic phase and the QGP phase at very high net baryon densities. The probability of multi-step baryonic collisions is high, thus leading to the formation of heavy baryonic resonance $N^{*}$ which subsequently decay to produce a $J/\psi$ state ($N^{*} \rightarrow J/\psi + N$). Fig.~\ref{fig:subthreshold_jpsi} shows the excitation of sub-threshold charm production in central Au+Au collisions. As evident from the results, the produced multiplicity of charmed hadrons does not exhibit any strong threshold behavior as seen in elementary p+p collisions. Due to secondary baryonic interactions in the model, essentially no sharp kinematic threshold exists anymore for hadron production. This results in the formation of $J/\psi$ mesons over a broad range of beam energies even below the elementary threshold. Below this energy, the heavy resonance which eventually decays into the charm hadrons cannot be produced in a first binary p+ p collision, but can be created in a secondary collision of two already excited states. Such secondary collisions between the excited states can reach a sufficiently high $\sqrt{s}$ to allow the production of a high mass resonance. Very recently near and below threshold multi-strange and charmed hadron production has been recalculated with the latest version (v4.0) of the UrQMD model. It incorporates density
and momentum dependent interaction potentials for all baryons, following a chiral mean-field (CMF) EoS which is consistent with neutron star observations and lattice QCD results~\cite{Steinheimer:2025trr}. For Au+Au collisions, sub-threshold production yield of $J/\psi$ mesons (along with other multi-strange and charmed hadrons) is found to exhibit a strong sensitivity to the underlying EoS of the produced medium as displayed in Fig.~\ref{fig:subthreshold_jpsi_new}. At 10A GeV beam energy and for $0 - 10 \%$ central collisions, the model predicts a $J/\psi$ multiplicity $5 (8) \times 10^{-6}$ when executed with the CMF EoS (cascade mode). Possible measurement of charmonium production as a function of beam energy in the SIS100 accelerator range thus serve as a novel probe of the QCD EoS at high baryon densities.

\begin{figure*}
    \centering
    \includegraphics[width=0.9\linewidth]{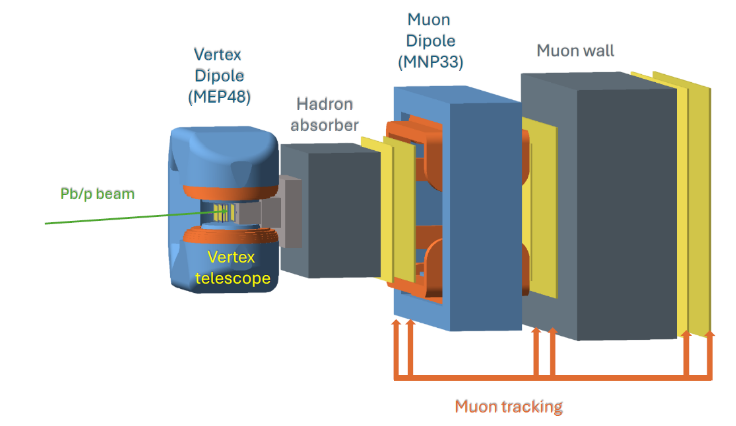}
    \caption{The foreseen setup of the NA60+/DiCE experiment at CERN SPS~\cite{NA60DiCE:2025abc}. The beam will impinge on the target from the left.}
    \label{fig:setup_na60_plus}
\end{figure*}

\section{Results of physics performance simulations}
\label{sec:4}
\begin{figure}
    \centering
     \includegraphics[width=1.0\linewidth]{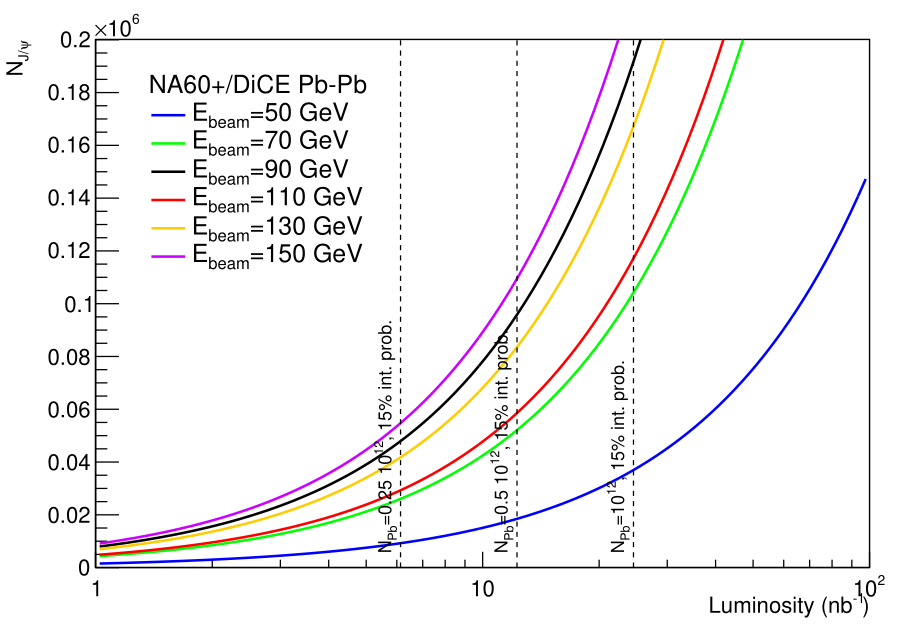}
        \caption{Foreseen statistics of $J/\psi$ signal measured in Pb+Pb collisions by the NA60+ spectrometer, as a function of the integrated beam luminosity ($L_{int}$), for different values of the incident beam energy. The figure is adopted from~\cite{NA60DiCE:2025abc}.}
    \label{jpsi_lumi_NA60_plus}
\end{figure}


\begin{figure*}
    \centering
     \includegraphics[width=0.45\linewidth]{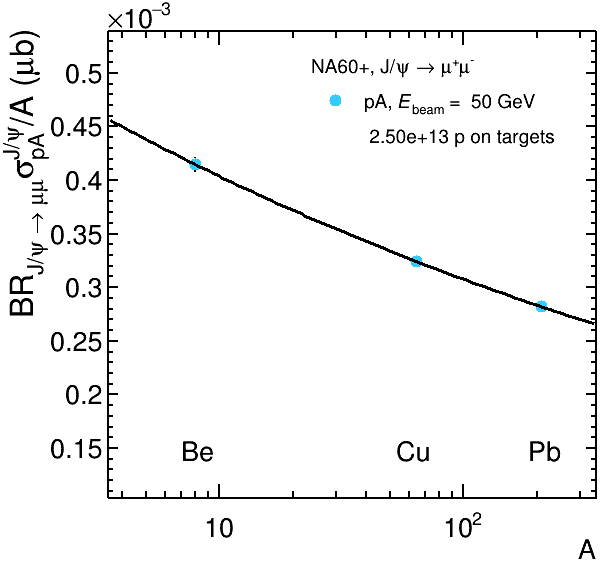}
    \includegraphics[width=0.45\linewidth]{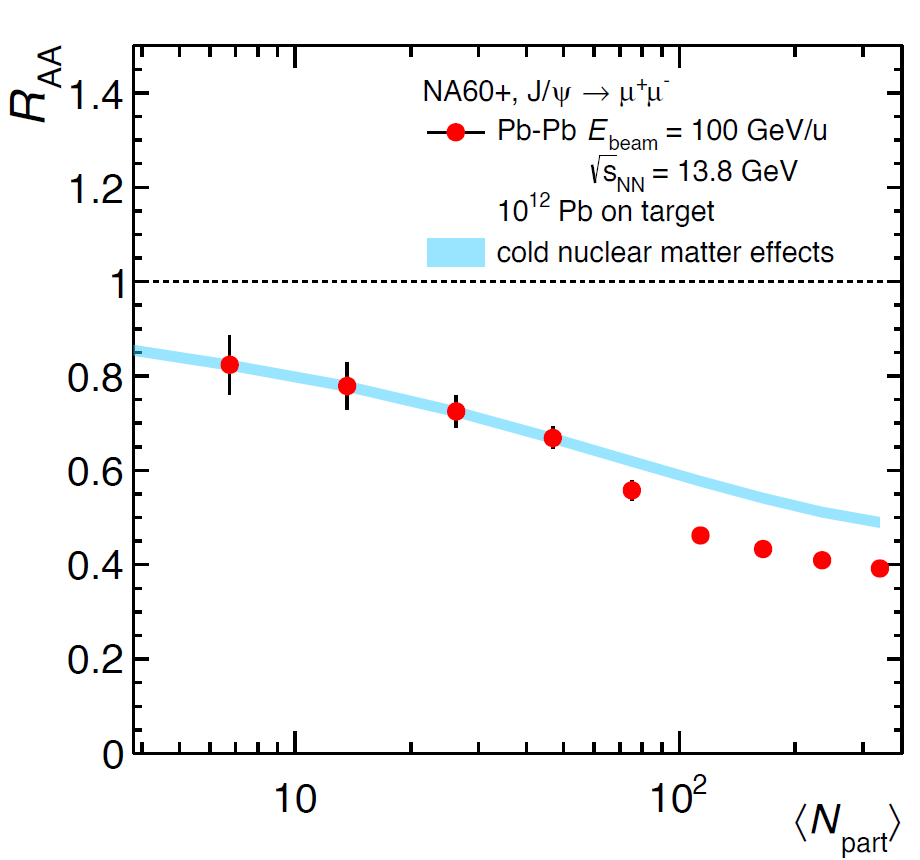}
    \caption{(left) Expected variation of the per nucleon $J/\psi$ production cross section as a function of mass number, in 50 GeV p+A collisions measured via the $\mu^{+}\mu^{-}$ decay channel for three nuclear targets. (right) Anticipated performance of the $R_{AA}$ measurement of $J/\psi$ production in 100 A GeV Pb+Pb collisions as a function of mean number of participant nucleons ($<N_{part}>$). A $30 \%$ anomalous suppression in addition to cold nuclear matter (CNM) effects is assumed for the five most central ($N_{part} > 80$) bins. The plots are adopted from~\cite{NA60DiCE:2025abc}.}
    \label{fig:results_NA60_plus}
\end{figure*}
 
 To realize the enormous physics potential of studying charmonium production in low energy collisions, different experimental programs are currently being designed to perform dedicated measurements. In this section we briefly discuss some of the major results of physics performance simulations for $J/\psi$ detection under realistic experimental conditions. The NA60+/DiCE ~\cite{Scomparin:2021xvy,Borysova:2022nyr,NA60_plus_LOI,Arnaldi:2023zlh,Scomparin:2023bbu,Alocco:2024hvm,Usai:2024row,NA60DiCE:2025qra,Arnaldi:2025ikz,NA60DiCE:2025abc} experiment plans to measure the $J/\psi$ production in p+A as well as in Pb+Pb collisions using the di-muon decay channel at various beam energies below 158 A GeV, that will be available during the data taking period. A schematic sketch of the foreseen detector setup optimized through simulation is shown in Fig.~\ref{fig:setup_na60_plus}~\cite{NA60DiCE:2025abc}. It has a close resemblance with the previous NA60 experiment that performed di-muon measurements in p+A and In+In collisions at top SPS energy. The experimental setup mainly comprises a vertex telescope (VT), to precisely measure the momentum and production angle of the charged particle tarcks and a muon spectrometer (MS) to measure the muon tracks which penetrate the thick hadron absorber placed downstream of the VT. To avoid degradation of muon kinematics due to multiple scattering and energy loss inside the hadron absorber, the reconstructed muon candidates in the MS will be matched in the coordinate and momentum space with the corresponding tracks in the VT. With increasing beam energy, the MS will be moved downstream to ensure a constant rapidity coverage around mid-rapidity, in the centre-of-mass system of the collisions. The thickness of the hadron absorber will be simultaneously increased to cope with the larger hadron multiplicity. The position of the VT will however remain unchanged due to its large angular acceptance. The target system is composed of five 1.5 mm thick Pb disks, for heavy-ion runs. For p+A collisions the system will be composed of multiple sub-targets of different nuclear species, simultaneously exposed to an incident proton beam. The use of Be, Cu and Pb targets is presently foreseen as default. The individual target thicknesses will be chosen to collect comparable event statistics 
for each nuclear species. Five identical silicon pixel stations positioned at $7< z< 38$ cm will comprise the VT. Each station of the telescope will be made of four large area monolithic pixel sensors of low material budget offering an intrinsic position resolution of 5 $\mu$m or better. The planes of the vertex spectrometer will be housed inside the C-shaped dipole magnet MEP48, originally constructed for the PS170 experiment and having a 400 mm high vertical gap. The magnet features in total 68 t of iron and 3 t of copper and offers a peak field of 1.47 T at a maximum current of 2000 A. The thick hadron absorber separates the muon spectrometer from the target region and filters out the produced hadrons. In addition to suppressing the hadronic background, the absorber also helps to shield the MS from the non-interacting beam particles and nuclear fragments produced in beam-target interactions and acts as a part of the
radiation protection of the experiment. Various materials like graphite (C), BeO, iron (Fe), concrete etc. of varying radiation length ($X_{0}$) and nuclear interaction length ($\lambda_{I}$) are being considered for the construction of the absorber. The three main elements of the absorber are the core and the plug (embedded in the core) located between VT and MS and the muon wall which precedes the last two stations of the MS that act as muon identifiers ($\mu$ID). In the currently considered proposal, the plug is of conical shape having the same length as the core of the absorber and made out of tungsten. The core has a modular structure with the first three modules made out of BeO (pre-absorber) followed by high density ($\rho \simeq 2$ gm/cm$^3$) carbon. With this design, the effective absorber thickness becomes $7\lambda_{I}$ in the so called low energy setup, at a beam energy of 20 A GeV  ($\sqrt{s_{NN}}= 6.3$ GeV) and $14\lambda_{I}$ in the long setup at a beam energy of 150 A GeV  ($\sqrt{s_{NN}} = 16.8$ GeV). Candidate muon tracks will be reconstructed in the MS, by means of six tracking stations having modular structure based on trapezoidal design. The two upstream stations are followed by MNP33 H-shaped dipole magnet and by two further tracking stations. The magnet features a useful aperture width of 2450 mm (full aperture 3100 mm) and a pole gap of 600 mm. The area of the MS stations corresponds to the opening of the MNP33 magnet. Particle hits measured before and after the magnet will help to determine the particle momentum in the MS. For efficient operation, the MNP33 magnet will be mounted on a rail system to enable movement along the $z$-direction, to adjust for different beam energies. A  3$\lambda_{I}$ thick “muon wall” will act as a final filter by reducing residual background by removing the punch through hadrons escaping the main absorber and the low-momentum muons. Two larger area additional chambers of coarser resolution, located behind the absorber wall will provide the required trigger for muon identification.


\begin{figure*}
    \centering
     \includegraphics[width=1.0\linewidth]{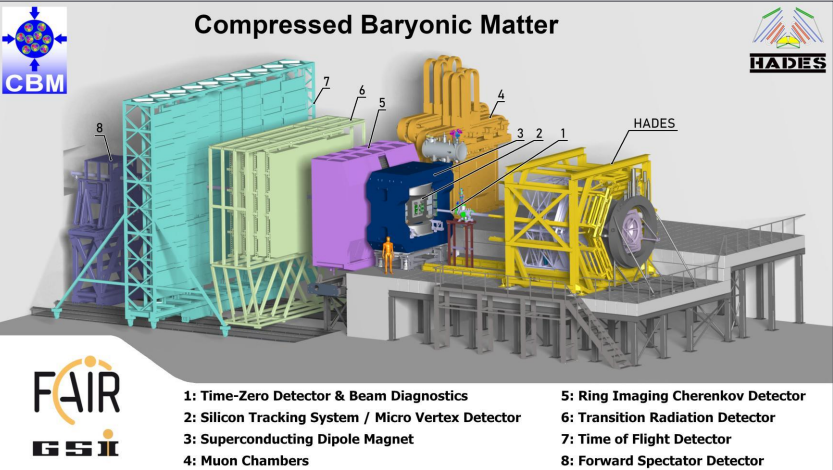}
        \caption{Schematic of the CBM experimental setup along with the various sub-detectors at FAIR SIS100 accelerator complex. The beam is entering from right.}
    \label{fig:setup_cbm}
\end{figure*}


With this foreseen detector setup detailed physics performance simulations have been carried out to evaluate the feasibility of measuring thermal di-muons and charm(onium) in low energy nuclear collisions.  
To identify the possible QGP effects on $J/\psi$ production, the NA60+ experiment aims to measure $R_{AA}$ as a function collision centrality at different collision energies. The corresponding production cross section in p+p collisions will be estimated by extrapolating the results from p+A collisions, collected by exposing various nuclear targets to the incident proton beams. The strength of CNM effects will be evaluated by analyzing the A-dependence of the $J/\psi$ production in p+A collisions and extrapolating the same to Pb+Pb collisions using standard Glauber framework as a function of collision centrality.  At each beam energy with typically $6 \times 10^{11}$ Pb ions on $15 \%  \lambda_{I}$ thick Pb target, significant $J/\psi$ statistics is expected to be collected down to a beam energy of 50 GeV per nucleon. The anticipated $J/\psi$ yield as a function of integrated luminosity ($L_{int}$) is displayed in Fig.~\ref{jpsi_lumi_NA60_plus}. 
As no measurement exists below 158 GeV to estimate nuclear absorption, a constant value of $\sigma_{abs}^{J/\psi} =7.6$ mb, as was extracted from 158 GeV p+A data by the NA60 experiment, is assumed in the beam energy domain 50 - 158 A GeV to model the suppression due to CNM effects. An additional 30 $\%$ suppression is hypothesized in $0 - 40 \%$ central collisions to include the effects of a deconfined medium. The vertical lines correspond to values of $L_{int}$ obtained with different numbers of Pb ions incident on the target. With the above assumptions of CNM effects and anomalous suppression, the per-nucleon $J/\psi$ production cross section in 50 GeV p+A collisions and the centrality dependence of $J/\psi$ $R_{AA}$ in 100 A GeV Pb+Pb collisions are shown in Fig.~\ref{fig:results_NA60_plus}. As evident from the results, with the foreseen statistics to be collected by the NA60+ setup, even a $20 \%$ level of anomalous suppression in most central Pb+Pb collisions can be clearly identified. 
As the melting of charmonium states inside a plasma depends on the binding energy of the resonance states, it will be interesting to measure the production of $\psi(2S)$ and $\chi_{c}$ states. However their detection at lower energies will be more difficult due to their extremely small yield. The identification of $\psi(2S)$ states via di-muon decay channel appears to be feasible down to 120 A GeV beam energy. Measurement of $\chi_{c} \rightarrow J/\psi + \gamma$ is also in principle possible by detecting the conversion $e^{+}e^{-}$ pair from the $\gamma$. A detailed detailed study of the feasibility of these measurements is in progress.\\

 On the other hand, the SIS-100 accelerator ring at the FAIR accelerator complex~\cite{Durante:2019hzd,Selyuzhenkov:2020djo,Reimann:2025pgc,Leifels:2025ucn} currently being commissioned in Darmstadt, Germany will deliver high intensity  beams of kinetic energy up to 30 GeV for protons, 14 A GeV for light nuclei (eg: Ca, Ni) and 11 A GeV for heavy nuclei (eg: Au, Pb). The physics program of the CBM experiment~\cite{CBM:2016kpk,Agarwal:2023otg,Hohne:2024kpd} at FAIR SIS100 thus includes the pioneering  measurement of subthreshold $J/\psi$ production in heavy-ion collisions below the kinematic production threshold. In addition CBM also plans to measure $J/\psi$ production in p+A collisions to probe the CNM effects on charm production at such low energies. Such measurements will be enabled by operating the CBM detector system at the highest ever interaction rate for heavy-ion collisions of 10 MHz. One unique feature of the CBM experiment is that it will perform $J/\psi$ measurements both via $\mu^{+}\mu^{-}$ and $e^{+}e^{-}$ decay channels. Fig.~\ref{fig:setup_cbm} shows the currently foreseen CBM fixed target experimental setup, including several sub detectors specially designed to cope with the requirements of high-rate data acquisition and precise event reconstruction in a high particle density environment~\cite{Teklishyn:2025fhl}. The silicon tracking system (STS) and the micro vertex detector (MVD) perform the tracking and vertex reconstruction with high spatial resolution. The STS, placed inside the 1 Tm superconducting dipole magnet, precisely determines the momentum of the charged particle tracks with a polar angle coverage of $2.5^{o} < \theta < 25^{o}$ and full $4\pi$ coverage in azimuth~\cite{Heuser:2024adt,CBM:2025voh,CBM:2025mnp}. Upstream of the STS, reconstruction of secondary decay vertices are performed by the MVD\footnote{The foreseen rate capability of MVD is 300 kHz and thus will not be used during high rate $J/\psi$ measurements.}. The Forward Spectator Detector (FSD), that records spectator fragments emitted at small forward angles determines the event geometry, by estimating  the collision centrality and the orientation of the reaction plane. Global tracking and particle identification are accomplished through several dedicated detector systems. For measuring dileptons, the Muon chamber (MuCh)  detector~\cite{Kumar:2021xpj,Agarwal:2025kle,Ghosh:2025tjz,Ghosh:2025hmc} system is employed for identifying the muon pairs produced in the collisions whereas the Ring Imaging Cherenkov (RICH) detector~\cite{Adamczewski-Musch:2017pri} and the Transition Radiation Detector (TRD)~\cite{Kahler:2020iln} are used for electron identification in complementary momentum ranges. Identification of charged hadrons based on precise timing measurements will be enabled by the multi gap resistive plate chamber (MRPC) based Time-of-Flight (ToF)~\cite{Deppner:2018kcf} wall together with the T0~\cite{Rost:2023sue} time reference detector. The design of the CBM setup has a modular structure and can be suitably reconfigured based on the physics goals of a particular measurement campaign. Each configuration comprises a set of dedicated detector subsystems, that enables the experiment to prioritize either hadron or lepton observables, or to be operated at maximum interaction rates for studying rare probes. Detection of $J/\psi$ mesons via the di-muon decay channel will be carried out by the CBM MUON setup~\cite{Senger:2021spm} comprising the sub detectors  STS, MuCh, TRD, ToF, and FSD. The MVD and RICH will not be included during the muon runs. The design of the MuCh system optimized for $J/\psi$ detection comprises four stations all made of gaseous detectors and five hadron absorbers. Each detector station is further made of three layers. Gas Electron Multiplier (GEM) detectors are used in the first two stations to handle the large particle rates and high level of accumulated radiation dose while resistive plate chamber (RPC) or straw tube tracker (STT) may be used in the third and fourth stations. The first absorber has a total thickness of 58 cm and is made of 30 cm concrete and 28 cm graphite ($\rho \simeq$ 1.78 gm/cm$^3$). The remaining absorbers are  made of iron to stop the hadrons produced during the collision. The second, third and fourth absorbers have thickness of 20 cm, 20 cm and 30 cm respectively and the final absorber is made of a 100 cm thick slab of iron to ensure that only high energy muons likely to come from $J/\psi$ decays, can reach till the end of the detector. Additionally TRD serves as a fifth detector to record the trigger hits for $J/\psi$ detection. \\
 
Extensive simulations are being carried out to study the performance of the CBM di-muon setup for the detection of sub-threshold $J/\psi$ production in 10 A GeV Au+Au collisions as well as $J/\psi$ production in 30 GeV p+Au collisions. For the anticipated sub-threshold $J/\psi$ multiplicity of $5 \times 10^{-6}$ as obtained with the UrQMD model, preliminary results clearly establish the feasibility of these measurements with reasonable signal statistics when operated at the highest interaction rate of 10 MHz~\cite{Sharma:2025ghp}. One advantage of the di-muon measurements with CBM is that the background due to DY process or correlated open charm cdecay will be negligible at SIS100 energies. A clear $J/\psi$ mass peak is visible over the background.


\section{Summary}
\label{sec:5}

 The study of hard probes is one of the key tools at our disposal from the very beginning of the quest for QGP in relativistic nuclear collisions. Charmonium is the only hard probe extensively measured so far in fixed target heavy-ion collisions. The early theoretical prediction of charmonium suppression as an experimentally viable signature of color deconﬁnement and QGP formation, followed by the nearly immediate availability of experimental results, has triggered an enormous amount of studies. $J/\psi$ suppression was first experimentally observed at SPS by the NA38 experiment in 200 A GeV O+U and S+U collisions. Soon it was realized that $J/\psi$ production gets considerably suppressed in p+A collisions as well due to the presence of the nuclear matter of the target nucleus and the most important issue in the interpretation of the heavy-ion data as a signal of QGP is related to identify the amount of suppression that can be assigned to ordinary nuclear processes. Only when a precise understanding of the magnitude of the normal suppression is reached, any claim for anomalous suppression induced by the hot and dense medium produced in nuclear collisions can be validated. A huge experimental effort was thus invested at SPS by the NA38, NA51, NA50 and NA60 experiments to study charmonium ($J/\psi$, $\psi(2S)$) production for a variety of systems starting from p+p and p+A to In+In and Pb+Pb collisions. The $J/\psi$ suppression observed in p+A collisions has been found to increase with the size of the nuclear target and is traditionally analyzed within the Glauber model, with a parameter $\sigma_{abs}^{eff}$ representing an effective absorption cross section and quantifying the overall magnitude of the observed CNM suppression. Compilation of $\sigma_{abs}^{eff}$ values extracted from different fixed target collisions exhibits a complex dependence on the $J/\psi$ kinematics. $\sigma_{abs}^{eff}$ has also been found to increase with decreasing collision energy, suggesting amplification of the CNM suppression at lower beam energies. The larger values of $\sigma_{abs}$ estimated for $\psi(2S)$ states are attributed to its larger size and weaker binding as compared to $J/\psi$. A variety of CNM effects have been identified in literature, whose complex interplay leads to the observed pattern of $J/\psi$ suppression in p+A collisions. However, so far we have not been able to unambiguously disentangle the different CNM effects, so as to obtain a description of the data in terms of "first principle" theories of the different CNM effects. The major effects studied so far include the modification of the parton densities inside the target nucleus in the initial stage prior to $c\bar{c}$ production in a hard partonic scattering and final state absorption of the evolving $c\bar{c}$ pairs due to inelastic collisions with the target nucleons. The magnitude of the final state absorption cross section ($\sigma_{abs}^{J/\psi}$) however is sensitive to the behavior of the parton densities inside the target. An enhanced target parton density in the initial stage will result in enhanced $c\bar{c}$ production, which will be more enhanced for larger nuclei. Thus, in order to obtain the value of $\sigma_{abs}^{eff}$ that reproduces the data, larger values of $\sigma_{abs}^{J/\psi}$ would be required in case of enhanced nuclear parton densities. $\sigma_{abs}^{J/\psi}$, as estimated in different phenomenological studies, is also seen to get stronger with decreasing beam energy. The heavy-ion data collected for 200 A GeV O+U, O+Cu and S+U collisions by the NA38 experiment and 158 A GeV In+In collisions by the NA60 experiment have been found to be compatible with the nuclear absorption scenario. Only $20 - 30 \%$ additional $J/\psi$ suppression is seen for the most central bins of Pb+Pb collisions, which in principle can be attributed to the formation a deconfined medium.  Till date there are no measurements of charmonium production in A+A collisions below 158 A GeV. \\

 The upcoming NA60+/DiCE experiment at CERN SPS and CBM experiment at FAIR SIS100 facilities will perform pioneering measurements of charmonium production in low energy heavy-ion collisions. Both the experimental programs are part of the worldwide endeavor to study the structure and properties of the QCD phase diagram in the region of large baryon densities. These studies require much smaller centre-of-mass energies as compared to the top energies at RHIC and LHC ion colliders. The NA60+ experiment will approximately cover the energy interval $6 < \sqrt{s_{NN}} < 17$ GeV, whereas the CBM experiment will access a lower energy domain, $2.9 <\sqrt{s_{NN}} < 4.9$ GeV complementing the NA60+ energy range. Together these two experiments will explore an overall interval of baryon chemical potential $220 < \mu_{B} < 800$ MeV of the QCD phase diagram. Both the experiments have dedicated plans to measure $J/\psi$ production in p+A and A+A collisions. At NA60+ $J/\psi$ mesons will be identified using a state-of-the-art muon spectrometer. $J/\psi$ data collected in p+A collisions will be useful for calibration of the magnitude of CNM effects at low beam energies and to measure the contribution of intrinsic charm to the observed charmonium production. On the other hand, a beam energy scan of $J/\psi$ production in heavy-ion collisions might help to identify the onset of anomalous suppression, which in turn will be useful in the experimental determination of the threshold temperature or energy density for QGP production in nuclear collisions. Since charmonium production due to regeneration is negligible in this energy domain the exact traces of suppression can be clearly identified. Model calculations incorporating various hadronic and partonic sources of $J/\psi$ dissociation predict distinguishably different suppression patterns in this energy domain. The maximum beam energy that will be delivered by FAIR SIS100 accelerator is below the kinematic threshold for $J/\psi$ production. However theoretical models predict sub-threshold charm production via multiple collisions between excited nucleons in a high baryon density medium. FAIR SIS100 energies are thus suitable for the first observation of sub-threshold $J/\psi$ production. Measurements of $J/\psi$ production in small bins of collision energy will also be useful to study the EoS of dense baryonic matter expected to be produced at these energies. Differential measurements of $J/\psi$ production in proton induced  collisions as a function of rapidity will be helpful for experimental determination of the resonance nucleon inelastic cross section, a quantity not well constrained by theoretical models. In summary, the upcoming measurements of charmonium production in low energy p+A and A+A collisions offer opportunities to address open questions that are unanswered by existing measurements. The major challenge is the extremely small charm production cross section that requires beams with extremely high intensities and radiation hard detectors with large rate capabilities to ensure a statistically significant measurement. Extensive simulations with realistic detector models are being carried out by both the NA60+ and CBM Collaborations to demonstrate the feasibility of detecting and studying charmonium production in this nearly unchartered energy regime.

 

%

\begin{acknowledgements}
The author would like to thank Sayak Chatterjee and Arun Kumar Yadav for stimulating discussions. 
\end{acknowledgements}

{\bf Data Availability Statement:} No Data associated in the manuscript.


\end{document}